\documentclass[twocolumn,superscriptaddress,pra,aps,10pt,longbibliography]{revtex4-1}
\usepackage{times}
\usepackage{amssymb}
\usepackage{color}
\usepackage{amsmath}
\usepackage{mathrsfs}
\usepackage{amsbsy}
\usepackage{amsthm}
\usepackage{graphicx}
\usepackage{tikz}
\usepackage[OT1]{fontenc}

\usepackage{bbm}
\usepackage{bm}
\usepackage{epsfig}
\usepackage{xfrac}
\usepackage{xcolor}
\usepackage{enumerate}
\usepackage[shortlabels]{enumitem}

\definecolor{myurlcolor}{rgb}{0,0,0.7}
\definecolor{myrefcolor}{rgb}{0.8,0,0}
\usepackage[unicode=true,pdfusetitle, bookmarks=true,bookmarksnumbered=false,bookmarksopen=false, breaklinks=false,pdfborder={0 0 0},backref=false,colorlinks=true, linkcolor=myrefcolor,citecolor=myurlcolor,urlcolor=myurlcolor]{hyperref}

\newcommand{\blackdiamond}[1][fill=black]{\tikz [x=0.75ex,y=0.75ex,line width=.1ex,line join=round, yshift=-0.285ex] \draw  [#1]  (0,.5) -- (.5,1) -- (1,.5) -- (.5,0) -- (0,.5) -- cycle;}%
\newcommand{\bdia}{{\blackdiamond}}
\newcommand{\btri}{{\blacktriangle}}

\usepackage{braket}
\let\vaccent=\v 
\renewcommand{\v}[1]{\ensuremath{\mathbf{#1}}} 
\newcommand{\abs}[1]{\left| #1 \right|} 
\newcommand{\norm}[1]{\left\| #1 \right\|} 
\newcommand{\trace}{\mathrm{Tr}}
\newcommand{\tl}{{\textsc{l}}}
\newcommand{\hl}{{\textsc{hl}}}
\newcommand{\sql}{{\textsc{sql}}}

\newcommand{\mS}{{\mathcal{S}}}

\newcommand{\mA}{{\mathcal{A}}}

\newcommand{\mZ}{{\mathcal{Z}}}
\newcommand{\mH}{{\mathcal{H}}}

\newcommand{\mU}{{\mathcal{U}}}
\newcommand{\mN}{{\mathcal{N}}}

\newcommand{\mR}{{\mathcal{R}}}
\newcommand{\frakF}{{\mathfrak{F}}}
\newcommand{\frakS}{{\mathfrak{S}}}
\newcommand{\frakP}{{\mathfrak{P}}}
\newcommand{\frakQ}{{\mathfrak{Q}}}

\newcommand{\mP}{{\mathcal{P}}}
\newcommand{\mD}{{\mathcal{D}}}
\newcommand{\mE}{{\mathcal{E}}}

\newcommand{\mT}{{\mathcal{T}}}
\newcommand{\id}{{\mathbbm{1}}}

\newcommand{\scrS}{{\mathscr{S}}}

\newcommand{\vK}{{\v{K}}}

\newcommand{\bR}{{\mathbb{R}}}
\newcommand{\bC}{{\mathbb{C}}}
\newcommand{\bH}{{\mathbb{H}}}
\newcommand{\bU}{{\mathbb{U}}}

\renewcommand{\Re}{{\mathrm{Re}}}

\newcommand{\ann}{{\hat a}}
\renewcommand{\epsilon}{\varepsilon}
\newcommand{\appropto}{\mathrel{\vcenter{
  \offinterlineskip\halign{\hfil$##$\cr
    \propto\cr\noalign{\kern2pt}\sim\cr\noalign{\kern-2pt}}}}}
\newcommand{\dket}[1]{\vert {#1} \rangle \! \rangle} 
\newcommand{\dbra}[1]{\langle \! \langle {#1} \vert} 
\newcommand{\dbraket}[1]{\langle \! \langle {#1} \rangle \! \rangle}
\let\baraccent=\= 
\renewcommand{\=}[1]{\stackrel{#1}{=}} 

\newcommand{\thmref}[1]{\hyperref[#1]{Theorem~\ref{#1}}}
\newcommand{\lemmaref}[1]{\hyperref[#1]{Lemma~\ref{#1}}}
\newcommand{\figref}[1]{\hyperref[#1]{Fig.~\ref{#1}}}
\newcommand{\tableref}[1]{\hyperref[#1]{Table~\ref{#1}}}
\newcommand{\figaref}[1]{\hyperref[#1]{Fig.~\ref{#1}a}}
\newcommand{\figbref}[1]{\hyperref[#1]{Fig.~\ref{#1}b}}
\newcommand{\figcref}[1]{\hyperref[#1]{Fig.~\ref{#1}c}}
\newcommand{\figdref}[1]{\hyperref[#1]{Fig.~\ref{#1}d}}
\newcommand{\figeref}[1]{\hyperref[#1]{Fig.~\ref{#1}e}}
\renewcommand{\eqref}[1]{\hyperref[#1]{Eq.~(\ref{#1})}}
\newcommand{\secref}[1]{\hyperref[#1]{Sec.~\ref{#1}}}
\newcommand{\secsref}[2]{\hyperref[#1]{Sec.~\ref{#1}-\ref{#2}}}
\newcommand{\eqsref}[2]{\hyperref[#1]{Eqs.~(\ref{#1})-(\ref{#2})}}
\newcommand{\appref}[1]{\hyperref[#1]{Appx.~\ref{#1}}}

\usepackage{dsfont}

\newtheorem{theorem}{Theorem}

\begin{document}

\title{Asymptotic theory of quantum channel estimation}

\author{Sisi Zhou}
\affiliation{Department of Physics, Yale University, New Haven, Connecticut 06511, USA}
\affiliation{Pritzker School of Molecular Engineering, The University of Chicago, Illinois 60637, USA}

\author{Liang Jiang}
\affiliation{Pritzker School of Molecular Engineering, The University of Chicago, Illinois 60637, USA}

\date{\today}

\begin{abstract}
The quantum Fisher information (QFI), as a function of quantum states, measures the amount of information that a quantum state carries about an unknown parameter. The (entanglement-assisted) QFI of a quantum channel is defined to be the maximum QFI of the output state assuming an entangled input state over a single probe and an ancilla. 
In quantum metrology, people are interested in calculating the QFI of $N$ identical copies of a quantum channel when $N \rightarrow \infty$, which is called the asymptotic QFI. Over the years, researchers found various types of upper bounds of the asymptotic QFI, but they were proven achievable only in several specific situations. It was known that the asymptotic QFI of an arbitrary quantum channel grows either linearly or quadratically with $N$. Here we show that a simple criterion can determine whether the scaling is linear or quadratic. In both cases, the asymptotic QFI and a quantum error correction protocol to achieve it are computable via a semidefinite program. When the scaling is quadratic, the Heisenberg limit, a feature of noiseless quantum channels, is recovered. When the scaling is linear, we show the asymptotic QFI is still in general larger than $N$ times the single-channel QFI and furthermore, sequential estimation strategies provide no advantage over parallel ones. 
\end{abstract}

\maketitle

\section{Introduction} 
\label{sec:intro}

Quantum metrology studies parameter estimation in a quantum system~\cite{giovannetti2011advances,degen2017quantum,braun2018quantum,pezze2018quantum,pirandola2018advances}. Usually, a quantum probe interacts with a physical system and the experimentalist performs measurements on the final probe state and infers the value of the unknown parameter(s) in the system from the measurement outcomes. It has wide applications in frequency spectroscopy~\cite{sanders1995optimal,bollinger1996optimal,huelga1997improvement,leibfried2004toward}, gravitational-wave detectors~\cite{caves1981quantum,yurke19862,berry2000optimal,higgins2007entanglement} and other high-precision measurements~\cite{buvzek1999optimal,gorecki2020pi,giovannetti2004quantum,valencia2004distant,de2005quantum}. 

The quantum Fisher information (QFI), 
which is inversely proportional to the minimum estimation variance, 
characterizes the amount of information a quantum state carries about an unknown parameter~\cite{helstrom1976quantum,holevo2011probabilistic,paris2009quantum,braunstein1994statistical}. To explore the fundamental limit on parameter estimation, we usually consider the situation where the number of quantum channels $N$ (or the probing time $t$) is large. The Heisenberg limit (HL), an $O(N^2)$ (or $O(t^2)$) scaling of the QFI, is the ultimate estimation limit allowed by quantum mechanics. It could be obtained, for example, using GHZ states in noiseless systems~\cite{giovannetti2006quantum,leibfried2004toward}. On the other hand, the standard quantum limit (SQL), an $O(N)$ (or $O(t)$) scaling of the QFI, usually appears in noisy systems and could be achieved using product states. Much work has been done towards determining whether or not the HL is achievable for a given quantum channel and some necessary conditions were derived~\cite{fujiwara2008fibre,ji2008parameter,escher2011general,hayashi2011comparison,demkowicz2012elusive,kolodynski2013efficient,knysh2014true,demkowicz2014using,sekatski2017quantum,demkowicz2017adaptive,zhou2018achieving,yuan2017fidelity,katariya2020geometric,yuan2017quantum}.

In general, the asymptotic QFI of a quantum system , i.e. the QFI in the $N \rightarrow \infty$ limit~\cite{kolodynski2013efficient}, follows either the HL or the SQL
and there was not a unified approach to determine the scaling. 
For quantum channels where the scalings are known, it is also crucial to understand how to achieve the asymptotic QFI. 
For example, for unitary channels, the HL is achievable and a GHZ state in the multipartite two-level systems consisting of the lowest and highest energy states is optimal~\cite{giovannetti2006quantum}. Under the effect of noise, a variety of quantum strategies were also proposed to enhance the QFI~\cite{caves1981quantum,wineland1992spin,huelga1997improvement,ulam2001spin,demkowicz2013fundamental,chaves2013noisy,gefen2016parameter,plenio2016sensing,albarelli2017ultimate,albarelli2019restoring,matsuzaki2011magnetic,chin2012quantum,smirne2016ultimate,liu2017quantum,xu2019transferable,chabuda2020tensor,zhou2019optimal}, but no conclusions for general quantum channels were drawn.
One natural question to ask is whether entanglement between probes can improve the QFI. For example, when estimating the noise parameter in the dissipative low-noise channels~\cite{hotta2005ancilla,hotta2006n} or teleportation-covariant channels~\cite{pirandola2017fundamental,pirandola2017ultimate,takeoka2016optimal,laurenza2018channel} (e.g. Pauli or erasure channels), the asymptotic QFI follows the SQL and is achievable using only product states. 
However, when estimating the phase parameter in dephasing channels, although the HL is still not achievable, product states are no longer optimal and the asymptotic QFI is then achievable using spin-squeezed states~\cite{huelga1997improvement,ulam2001spin,demkowicz2014using}. 

Given a quantum channel, we aim to answer the following two important questions: how to determine whether the HL is achievable, and in both cases, how to find a metrological protocol achieving the asymptotic QFI? In this paper, we answer these two open problems in the setting of entanglement-assisted channel estimation by providing an optimal quantum error correction (QEC) metrological protocol which entangles both the probe and a clean ancillary system. QEC has been a powerful tool widely used in quantum computing and quantum communication to protect quantum information from noise~\cite{lidar2013quantum,knill1997theory,gottesman2010introduction,dur2007entanglement}. In quantum metrology, QEC is also useful in protecting quantum signal from quantum noise~\cite{kessler2014quantum,arrad2014increasing,dur2014improved,ozeri2013heisenberg,unden2016quantum,reiter2017dissipative,sekatski2017quantum,lu2015robust,demkowicz2017adaptive,zhou2018achieving,layden2018spatial,layden2019ancilla,gorecki2019quantum,tan2019quantum,kapourniotis2019fault,zhuangi2019distributed,chen2020fluctuationenhanced}. Here is a typical example: 
when a qubit is subject to a Pauli-$Z$ 
signal and a 
Pauli-$X$
noise, the QFI follows the SQL if no quantum control is added, but the HL is recoverable using fast and frequent QEC~\cite{kessler2014quantum,arrad2014increasing,dur2014improved,ozeri2013heisenberg,unden2016quantum,reiter2017dissipative}. The result could be generalized to any system with a signal Hamiltonian and Markovian noise~\cite{demkowicz2017adaptive,zhou2018achieving}. These QEC protocols, however, can only estimate Hamiltonian parameters and all rely on fast and frequent quantum operations which have limited practical applications. 

In this paper, 
we construct a two-dimensional QEC protocol which reduces every quantum channel to a single-qubit dephasing channel where both the phase and the noise parameter could vary w.r.t. the unknown parameter. 
We first identify the asymptotic QFI for all single-qubit dephasing channels (where the unknown parameter is encoded in both the noise and phase parameters) and then show that the asymptotic QFI of the logical dephasing channel is no smaller than the one of the original quantum channel after optimizing over the encoding and the recovery channel, proving the sufficiency of our QEC protocol. 
Using the above proof strategy, we obtain the asymptotic theory of quantum channel estimation, 
closing a long-standing open question in theoretical quantum metrology. We also push one step further towards achieving the ultimate estimation limit in practical quantum sensing experiments by providing efficiently computable asymptotic QFIs and corresponding optimal estimation protocols.

\section{Overview}

\subsection{Preliminaries}

The quantum Cram\'{e}r-Rao bound is a lower bound of the estimation precision~\cite{helstrom1976quantum,holevo2011probabilistic,paris2009quantum,braunstein1994statistical}, 
\begin{equation}
\delta \omega \geq \frac{1}{\sqrt{N_{\rm expr} F(\rho_\omega)}},
\end{equation} 
where $\omega$ is an unknown real parameter to be estimated, $\delta \omega$ is the standard deviation of any unbiased estimator of $\omega$, $N_{\rm expr}$ is the number of repeated experiments and $F(\rho_\omega)$ is the QFI of the state $\rho_\omega$. The quantum Cram\'{e}r-Rao bound is saturable asymptotically ($N_{\rm expr} \gg 1$) using maximum likelihood estimators~\cite{casella2002statistical,lehmann2006theory}. Therefore, the QFI is a good measure of the amount of information a quantum state $\rho_\omega$ carries about an unknown parameter. It is defined by $F(\rho_\omega) = \trace\big( L^2 \rho_\omega\big)$,
where $L$ is a Hermitian operator called the symmetric logarithmic derivative (SLD) satisfying 
\begin{equation}
\dot\rho_\omega = \frac{1}{2}(\rho_\omega L + L \rho_\omega),
\end{equation} 
where $\dot{\star}$ denotes $\frac{\partial\star}{\partial\omega}$. We will use $L_A[B]$ to represent Hermitian operators satisfying $B = \frac{1}{2}(LA+AL)$. Here $L = L_{\rho_\omega}[\dot\rho_\omega]$. The QFI could also be equivalently defined through purification~\cite{fujiwara2008fibre,kolodynski2013efficient}:
\begin{equation}
\label{eq:purification}
F(\rho_\omega) = 4 \min_{\ket{\psi_\omega}:\trace_E(\ket{\psi_\omega}\bra{\psi_\omega}) = \rho_\omega} \braket{\dot\psi_\omega|\dot\psi_\omega},
\end{equation}
where $\rho_\omega \in \frakS(\mH_\mP)$, $\ket{\psi_\omega} \in \frakS(\mH_\mP \otimes \mH_E)$, $\mH_\mP$ is the probe system which we assume to be finite-dimensional, $\mH_E$ is an arbitrarily large environment and $\frakS(\star)$ denotes the set of density operators in $\star$.  

\begin{figure}[t]
\includegraphics[width=7cm]{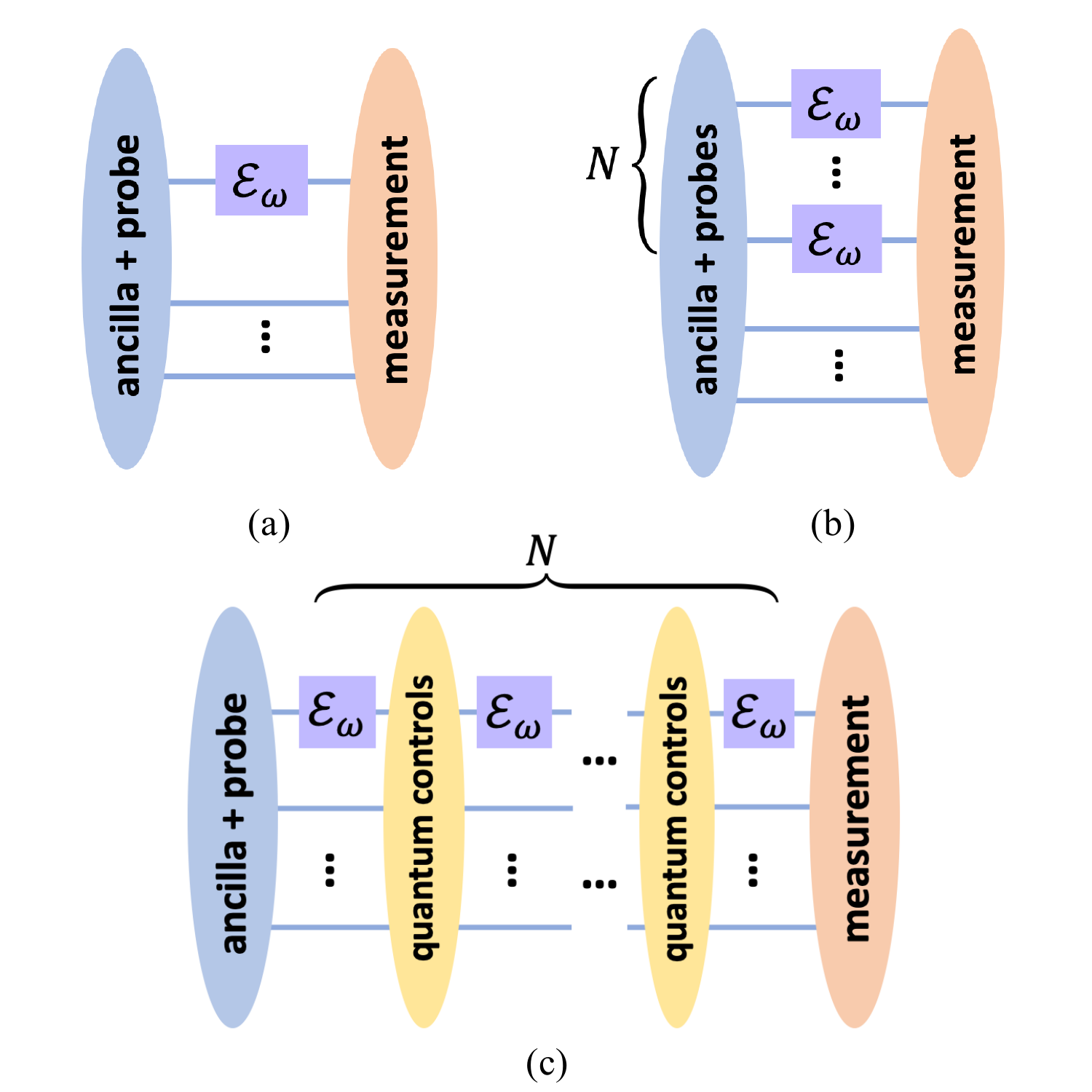}
\caption{\label{fig:protocol} 
(a) The single-channel QFI $\frakF_1(\mE_\omega) = \max_\rho F((\mE_\omega \otimes \id)(\rho))$. The ancillary system is assumed to be arbitrarily large. 
(b) Parallel strategies. $\frakF_N(\mE_\omega) = \frakF_1(\mE_\omega^{\otimes N}) = \max_\rho F((\mE_\omega^{\otimes N} \otimes \id)(\rho))$ for $N$ identical copies of $\mE_\omega$. (c) Sequential strategies. 
Let $F_N(\mE_\omega,\scrS)$ be the QFI of the output state, given a sequential strategy $\scrS$ which contains both an input state and quantum controls acting between $\mE_\omega$. $\frakF_N^{(\rm seq)}(\mE_\omega) = \max_{\scrS}F_N(\mE_\omega,\scrS)$ is the optimal QFI maximized over all sequential strategies.  $\frakF_N^{(\rm seq)}(\mE_\omega) \geq \frakF_N(\mE_\omega)$. 
} 
\end{figure}

We consider a quantum channel $\mE_\omega(\rho) = \sum_{i=1}^r K_i \rho K_i^\dagger$, where $r$ is the rank of the channel.
The entanglement-assisted QFI of $\mE_\omega$ (see \figaref{fig:protocol}) is defined by~\cite{fujiwara2008fibre,kolodynski2013efficient}, 
\begin{equation}
\label{eq:single}
\frakF_1(\mE_\omega) := \max_{\rho\in\frakS(\mH_\mP\otimes\mH_\mA)} F((\mE_\omega \otimes \id)(\rho)).
\end{equation}
Here we utilize the entanglement between the probe and an arbitrarily large ancillary system $\mH_\mA$. We will omit the word ``entanglement-assisted'' or ``ancilla-assisted'' in the definitions below for simplicity. Practically, the ancilla should be a quantum system with a long coherence time, e.g. nuclear spins~\cite{unden2016quantum} or any QEC-protected system~\cite{zhou2018achieving}. The ancilla also helps simplify the complicated calculation of the QFI. The convexity of QFI implies the optimal input state is always pure. Using the purification-based definition of the QFI (\eqref{eq:purification}), we have~\cite{fujiwara2008fibre,demkowicz2012elusive,kolodynski2013efficient}
\begin{align}
\label{eq:one-shot-2}
\frakF_1(\mE_\omega) 
&= 4 \max_{\rho\in\frakS(\mH_\mP)} \min_{\substack{\vK' = u\vK\\\forall u,\text{~s.t.~} u^\dagger u = I}} \trace(\rho \dot\vK'^\dagger \dot\vK')\\
\label{eq:one-shot}
&= 4 \min_{\substack{\vK' = u\vK\\\forall u,\text{~s.t.~} u^\dagger u = I}} \|\dot\vK'^\dagger \dot\vK'\| = 4 \min_{h \in \bH_r} \norm{\alpha}
,
\end{align}
where $\norm{\cdot}$ is the operator norm, $\bH_r$ is the space of $r\times r$ Hermitian matrices and $\vK = (K_1,\ldots,K_r)^T$. $\vK' = (K'_1,\ldots,K'_r)^T = u\vK$ represents all possible Kraus representations of $\mE_\omega$ via isometric transformations $u$~\cite{fujiwara2008fibre}. Let $h = i u^\dagger \dot u$ and $\alpha = \dot\vK'^\dagger \dot\vK' = (\dot\vK - ih\vK)^\dagger (\dot\vK-ih\vK)$. The minimization is performed over arbitrary Hermitian operator $h$ in $\bC^{r\times r}$~\cite{demkowicz2012elusive}. 
Any purification of the optimal $\rho$ in \eqref{eq:one-shot-2} is an optimal input state in $\mH_\mP\otimes \mH_\mA$ and it in general depends on the true value of $\omega$ and should be chosen adaptively throughout the experiment~\cite{barndorff2000fisher,gill2000state}. The QFI $\frakF_1(\mE_\omega)$ can be found via a semidefinite program (SDP)~\cite{demkowicz2012elusive,kolodynski2013efficient}, as well as the optimal input state (see~\appref{app:algo}). As an example, we show in \appref{app:interfer} that when viewing the ancilla as a lossless arm in the Mach-Zehnder interferometer~\cite{huver2008entangled,demkowicz2009quantum}, the SDP in \appref{app:algo} leads to an SDP solving for the optimal input state with a definite photon number.

Consider $N$ identical copies of the quantum channel $\mE_\omega$~\cite{fujiwara2008fibre,demkowicz2012elusive} (see \figbref{fig:protocol}), let 
\begin{equation}
\frakF_N(\mE_\omega) := \frakF_1(\mE_\omega^{\otimes N}) = \max_\rho F((\mE_\omega^{\otimes N} \otimes \id)(\rho)).
\end{equation}
Clearly $\frakF_N \geq N \frakF_1$ using the additivity of the QFI. An upper bound on $\frakF_N(\mE_\omega)$ could be derived from \eqref{eq:one-shot}~\cite{fujiwara2008fibre,demkowicz2012elusive} (see also \appref{app:upper}), 
\begin{equation}
\label{eq:upper}
\frakF_N(\mE_\omega) \leq 4 \min_{h} \big( N \norm{\alpha} + N(N-1) \norm{\beta}^2 \big),
\end{equation}
where $\beta = i \vK^\dagger (\dot\vK - ih\vK)$. 
If there is an $h$ such that $\beta = 0$,
\begin{equation}
\label{eq:upper-2}
\frakF_N(\mE_\omega) \leq 4 \min_{h:\beta = 0}  N \norm{\alpha},
\end{equation}
and $\frakF_N(\mE_\omega)$ follows the SQL asymptotically.

{
The metrological protocols we considered in \figbref{fig:protocol} are usually called parallel strategies where $N$ identical quantum channels act in parallel on a quantum state~\cite{demkowicz2014using}. Researchers also consider sequential strategies where we allow quantum controls (arbitrary quantum operations) between each quantum channels (see \figcref{fig:protocol}). The QFI optimized over all possible inputs and quantum controls has the upper bound~\cite{demkowicz2014using,sekatski2017quantum},
\begin{multline}
\frakF^{\rm (seq)}_N(\mE_\omega) \leq 4 \min_{h} \Big( N \norm{\alpha} + \\ N(N-1) \norm{\beta}(\norm{\beta} + 2\sqrt{\norm{\alpha}})\Big).
\end{multline}
Therefore, $\forall h, \beta \neq 0$ is also a necessary condition to achieve the HL for sequential strategies. When the condition is violated, there exists an $h$ such that $\beta = 0$ and $\frakF^{\rm (seq)}_N(\mE_\omega)$ has the same upper bound (\eqref{eq:upper-2}) as $\frakF_N(\mE_\omega)$. Sequential strategies are more powerful than parallel strategies because they can simulate parallel strategies using the same input states and swap operators as quantum controls~\cite{demkowicz2014using}. 
}

\subsection{Main results}

{
In fact, the condition that $\forall h$, $\beta \neq 0$ is equivalent to $H \notin \mS$, where 
\begin{equation}
H= i \vK^\dagger \dot\vK,\quad \mS= {\rm span}_{\bH}\{K_i^\dagger K_j,\forall i,j\}.
\end{equation} 
Here ${\rm span}_{\bH}\{\cdot\}$ represents all Hermitian operators which are linear combinations of operators in $\{\cdot\}$. 
We call it the HNKS condition, or simply HNKS, an acronym for ``Hamiltonian-not-in-Kraus-span''. One can check that $H$ and $\beta$ are always Hermitian by taking the derivative of $\vK^\dagger \vK = I$. Note that different Kraus representations may lead to different $H$, but they only differ by some operator inside $\mS$, so whether $H \in \mS$ or  $H \notin \mS$ does not depend on the choice of Kraus representation. For a unitary channel $r = 1$ and $K_1 = U_\omega = e^{-iH\omega}$, $H = iU_\omega^\dagger \dot U_\omega$ is exactly the Hamiltonian for $\omega$, explaining its name. The HL is achievable for unitary channels because $\mS = {\rm span}_{\bH}\{I\}$ and we always have $H \notin \mS$ for nontrivial $H$. It is only possible to achievable the HL if HNKS holds. }

We will show in \secref{sec:HL} that HNKS 
is also a sufficient condition to achieve the HL for parallel strategies in \figbref{fig:protocol}, and hence, sequential strategies in \figcref{fig:protocol} that contain the former. We summarize this by the following theorem:
\begin{theorem}
\label{thm:HNKS}
$\frakF_N(\mE_\omega) = \Theta(N^2)$ if and only if $H\notin \mS$. Otherwise, $\frakF_N(\mE_\omega) = \Theta(N)$.  The statement is also true for $\frakF^{(\rm seq)}_N(\mE_\omega)$. 
\end{theorem}

Furthermore, 
the QFI upper bound in \eqref{eq:upper-2} is achievable asymptotically when $H \in \mS$ for both parallel and sequential strategies: 

\begin{theorem}
\label{thm:SQL}
When $H \in \mS$, 
\begin{equation}
\frakF_\sql(\mE_\omega) := \lim_{N\rightarrow \infty} \frakF_N(\mE_\omega)/N = 4 \min_{h:\beta = 0}\norm{\alpha}. 
\end{equation}
For any $\eta > 0$, there exists an input state $\ket{\psi_{\eta,N}}$ computable via an SDP such that $\lim_{N\rightarrow \infty} F((\mE_\omega^{\otimes N} \otimes \id)(\ket{\psi_{\eta,N}}))/N > \frakF_\sql(\mE_\omega) - \eta$. Furthermore, $\frakF_\sql^{(\rm seq)}(\mE_\omega) = \frakF_\sql(\mE_\omega)$.  
\end{theorem}

Note that $\frakF_\sql(\mE_\omega)$ is named ``asymptotic channel QFI'' in~\cite{kolodynski2013efficient}. 
The quadratic term of the QFI upper bound in \eqref{eq:upper} is also achievable  when $H \notin \mS$ for parallel strategies:

\begin{theorem}
\label{thm:HL}
When $H \notin \mS$, 
\begin{equation}
\frakF_\hl(\mE_\omega) :=  \lim_{N\rightarrow \infty} \frakF_N(\mE_\omega)/N^2 = 4 \min_{h}\norm{\beta}^2. 
\end{equation}
There exists an input state $\ket{\psi_{N}}$ computable via an SDP such that 
$F((\mE_\omega^{\otimes N} \otimes \id)(\ket{\psi_{N}}))/N^2 = \frakF_\hl(\mE_\omega)$. 
\end{theorem}

Note that without the help of the ancilla system, the QFI upper bound in \eqref{eq:upper-2} may not be  achievable asymptotically~\cite{knysh2014true,layden2019ancilla}. For example, the upper bound in \eqref{eq:upper-2} for phase estimation in amplitude damping channels is reduced by a factor of four without ancilla~\cite{knysh2014true,demkowicz2014using}. 

Although the theorems above for general quantum channel estimation were not proven before this work, we note here that in the special case of Hamiltonian estimation under Markovian noise with the assistance of fast and frequency quantum controls, i.e.~sequential strategies (\figcref{fig:protocol}) where $\mE_{\omega,dt}(\rho) = -i[\omega H',\rho]dt + \sum_i(L_i\rho L_i^\dagger - \frac{1}{2}\{L_i^\dagger L_i, \rho\}) + O(dt^2)$, $t$ is the probing time and $N=t/dt$, the results (when taking the limit $dt \rightarrow 0$) were already known. The necessity part was proven in~\cite{sekatski2017quantum,demkowicz2017adaptive,zhou2018achieving} and the sufficiency part was proven in~\cite{zhou2018achieving,zhou2019optimal}. In particular, it was shown that in this case, the HNKS condition reduces to the condition $H' \notin \mS'$ where $\mS' = {\rm span}_\bH\{I,L_i,L_i^\dagger,L_i^\dagger L_j,\forall i,j\}$ is called the Lindblad span. The condition was first named the ``Hamiltonian-not-in-Lindblad-span'' (HNLS) condition in~\cite{zhou2018achieving}. 

\thmref{thm:SQL} indicates that when HNKS is violated (which almost surely happens statistically), there is no advantage of sequential strategies over parallel strategies asymptotically, as conjectured in~\cite{demkowicz2014using}. Interestingly, similar results were discovered quantum channel discrimination, a related field~\cite{hayashi2002two,yuan2017fidelity,pirandola2019fundamental,chen2019zero,katariya2020geometric,chiribella2008memory,yang2019memory}. It was recently proven that sequential strategies cannot outperform parallel strategies asymptotically in asymmetric discrimination of two arbitrary quantum channels~\cite{hayashi2009discrimination,cooney2016strong,berta2020amortized,wang2019resource,fang2019chain}. Our result is different, however, because the QFI cannot be characterized as the limit of quantum relative entropy~\cite{hayashi2002two} and it is also unclear how to interpret the HNKS condition in terms of asymmetric channel discrimination. 
Moreover, we provide a constructive proof with explicit and efficiently computable QEC metrological protocols,  
which paves the way for practical implementation of error-corrected sensing schemes.

Based on the previous discussion, in order to prove the theorems, it is sufficient to provide a QEC protocol using parallel strategies which achieves the QFI upper bound (\eqref{eq:upper}) asymptotically both when $H \in \mS$ or $H \notin \mS$. Thus we will focus only on parallel strategies in the following. We first show \thmref{thm:HL} and \thmref{thm:SQL} are true for the generalized single-qubit dephasing channels in \secref{sec:dephasing} where both the phase and the noise parameter vary w.r.t. $\omega$. Then we will generalized the results to arbitrary quantum channels $\mE_\omega$ using a QEC protocol in \secsref{sec:QEC}{sec:SQL}. The two steps are summarized in \figref{fig:sketch}.

\section{Single-qubit dephasing channels} 
\label{sec:dephasing}

According to \eqref{eq:upper}, $\frakF_\hl \leq \frakF^{(u)}_\hl$ and $\frakF_\sql \leq \frakF^{(u)}_\sql$, where $\frakF_\hl^{(u)} := 4 \min_{h}\norm{\beta}^2$ and $\frakF_\sql^{(u)} := 4 \min_{h:\beta = 0}\norm{\alpha}$. {$^{(u)}$ refers to the upper bounds here.} In this section, we will show the above equalities hold for any single-qubit dephasing channel
\begin{equation}
\mD_\omega(\rho) = (1 - p) e^{-\frac{i \phi}{2} \sigma_z} \rho e^{\frac{i \phi}{2} \sigma_z} + p \sigma_z e^{-\frac{i \phi}{2} \sigma_z} \rho e^{\frac{i \phi}{2} \sigma_z} \sigma_z,
\end{equation}
which is the composition of the conventional dephasing channel $\rho \mapsto (1-p) \rho + p \sigma_z \rho \sigma_z$ ($0 \leq p < 1$) and the rotation in the $z$-direction $\rho \mapsto e^{-\frac{i \phi}{2} \sigma_z} \rho e^{\frac{i \phi}{2} \sigma_z}$. Both $p$ and $\phi$ are functions of an unknown parameter $\omega$. 
As shown in \appref{app:dephasing-upper}, the HNKS condition is equivalent to $p = 0$ and the QFI upper bounds for $\mD_\omega$ are
\begin{equation}
\label{eq:dephasing-bounds}
\frakF_{\hl}^{(u)}(\mD_\omega) = |\dot\xi|^2, \quad \frakF_{\sql}^{(u)}(\mD_\omega) = \frac{|\dot\xi|^2}{1 - |\xi|^2}, 
\end{equation}
where $\xi = \bra{0}\mD_\omega(\ket{0}\bra{1})\ket{1} = (1-2p)e^{-i\phi}$. 

Now we show that $\frakF_{\hl,\sql}(\mD_\omega) = \frakF_{\hl,\sql}^{(u)}(\mD_\omega)$ and provide the optimal input states in both cases. When HNKS is satisfied ($p = 0$), $\mD_\omega$ is unitary. Using the GHZ state $\ket{\psi_0} = \frac{1}{\sqrt{2}}\big(\ket{0}^{\otimes N} + \ket{1}^{\otimes N} \big)$ as the input state, we could achieve 
\begin{equation}
F(\mD_\omega^{\otimes N}(\ket{\psi_0}\bra{\psi_0})) = |\dot\xi|^2 N^2, 
\end{equation}
which implies $\frakF_{\textsc{hl}}(\mD_\omega) = \frakF^{(u)}_{\textsc{hl}}(\mD_\omega)$.

To calculate the optimal QFI when HNKS is violated ($p > 0$), we will use the following two useful formulae. For any pure state input $\ket{\psi_0}$ and output $\rho_\omega = \mD_\omega^{\otimes N}(\ket{\psi_0}\bra{\psi_0})$, we have, for all $N$, 
\begin{equation}
\label{eq:dephasing-sum}
F(\rho_\omega) = F_{p}(\rho_\omega) + F_{\phi}(\rho_\omega),
\end{equation}
where $F_{p}(\rho_\omega) = \trace(L_{p}^2 \rho_\omega)$ is the QFI w.r.t. $\omega$ when only the noise parameter $p$ varies w.r.t. $\omega$, where the SLD $L_{p}$ satisfies $\frac{\partial \rho_\omega}{\partial p}{\dot p} = \frac{1}{2}(L_{p} \rho_\omega + \rho_\omega L_{p})$. Similarly, $F_{\phi}(\rho_\omega)$ is the QFI w.r.t. $\omega$ when only the phase parameter $\phi$ varies w.r.t. $\omega$. The proof of \eqref{eq:dephasing-sum} is provided in \appref{app:dephasing-sum}. Another useful formula is the error propagation formula~\cite{pezze2009entanglement}, 
\begin{equation}
\label{eq:dephasing-var}
F(\rho) \geq \frac{1}{\braket{\Delta J^2}_{\rho}}\left( \frac{\partial \braket{J}_{\rho}}{\partial \omega}\right)^2,
\end{equation}
for arbitrary $\rho$ as a function of $\omega$ and arbitrary Hermitian operator $J$ where $\braket{J}_{\rho} = \trace(J \rho)$ and $\braket{\Delta J^2}_{\rho} = \braket{J^2}_{\rho} - \braket{J}_{\rho}^2$. The equality holds when $J$ is equal to the SLD operator of $\rho$.

\begin{figure}[tb]
\includegraphics[width=0.48\textwidth]{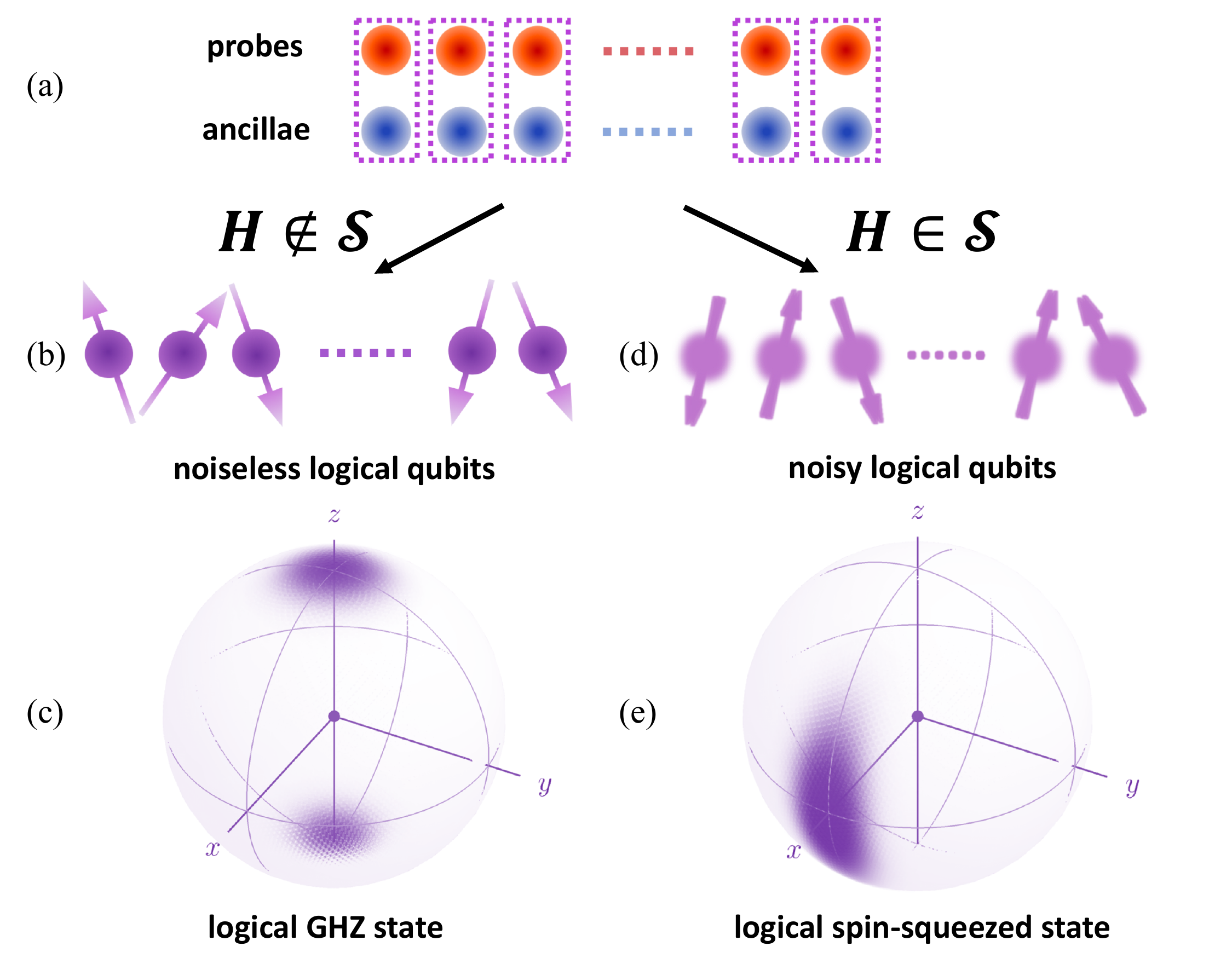}
\caption{\label{fig:sketch} 
The optimal metrological protocol. (a) The original physical system where we have $N$ noisy probes and $N$ noiseless ancillas. Each pair of probe-ancilla subsystem (purple box) encodes a logical qubit (see \secref{sec:QEC}). 
(b,c) When $H\notin \mS$, the logical qubits are noiseless. We choose the GHZ state of $N$-logical qubits as the optimal input.
(d,e) When $H \in S$, each logical qubit is subject to an effective dephasing noise. We choose the spin-squeezed state of the $N$-logical qubits with suitable parameters as the optimal input.
We plot the quasiprobability distribution $Q(\theta,\varphi) = \abs{\braket{\theta,\varphi|\psi}}^2$ on a sphere using coordinates $(x,y,z) = (\sin\theta\cos\varphi,\sin\theta\sin\varphi,\cos\theta)$~\cite{kitagawa1993squeezed}, where $\ket{\theta,\varphi} = (\cos\frac{\theta}{2}\ket{0} + e^{i\varphi}\sin\frac{\theta}{2}\ket{1})^{\otimes N}$ and $N = 50$. (Darker colors indicate larger values.) 
} 
\end{figure}

Consider an $N$-qubit spin-squeezed state~\cite{kitagawa1993squeezed,ulam2001spin}: 
\begin{equation}
\ket{\psi_{\mu,\nu}} = e^{- i\nu J_x}e^{-\frac{i \mu}{2}J_z^2} e^{-i\frac{\pi}{2}J_y} \ket{0}^{\otimes N},
\end{equation}
where $J_{x,y,z} = \frac{1}{2}\sum_{k=1}^N \sigma_{x,y,z}^{(k)}$ with $^{(k)}$ denote operators on the $k$-th qubit. 
Let $\ket{\psi_0} = e^{i\phi J_z}\ket{\psi_{\mu,\nu}}$. Using \eqref{eq:dephasing-sum} and \eqref{eq:dephasing-var}, 
we have for $\rho_\omega = \mD_\omega^{\otimes N}(\ket{\psi_0}\bra{{\psi_0}})$,
\begin{multline}
F(\rho_\omega) \geq \frac{1}{\braket{\Delta J^2_x}_{\rho_\omega}}\bigg(\frac{\partial \braket{J_x}_{\rho_\omega}}{\partial p} \dot p\bigg)^2 \\ + \frac{1}{\braket{\Delta J_y^2}_{\rho_\omega}}\bigg(\frac{\partial \braket{J_y}_{\rho_\omega}}{\partial \phi} \dot \phi\bigg)^2. 
\end{multline}
As shown in \appref{app:dephasing-squeezed}, as $N\rightarrow \infty$, with suitable choices of $(\mu,\nu)$, we have (up to the lowest order of $N$), $\braket{\Delta J_x^2}_{\rho_\omega} \approx \braket{\Delta J_y}^2_{\rho_\omega} \approx p(1-p)N$, $\frac{\partial{\braket{J_x}_{\rho_\omega}}}{\partial p} \dot p \approx -\dot p N$ and $\frac{\partial{\braket{J_y}_{\rho_\omega}}}{\partial \phi} \dot \phi \approx (1-2p)\dot\phi N/2$. For example, we can choose $\mu = 4 (\frac{2}{N})^{5/6}$ and $\nu = \frac{\pi}{2} - \frac{1}{2}\arctan{\frac{4\sin\frac{\mu}{2}\cos^{N-2}\frac{\mu}{2}}{1-\cos^{N-2}\mu}}$. The corresponding $\ket{\psi_{\mu,\nu}}$ is illustrated in \figeref{fig:sketch} using the quasiprobability distribution $Q(\theta,\varphi) = \abs{\braket{\theta,\varphi|\psi_{\mu,\nu}}}^2$ on a sphere~\cite{kitagawa1993squeezed}. Therefore, 
\begin{equation}
F(\rho_\omega) \geq 
\frac{|\dot\xi|^2}{1 - |\xi|^2} N+ o(N),
\end{equation}
which implies $\frakF_{\textsc{sql}}(\mD_\omega) = \frakF^{(u)}_{\textsc{sql}}(\mD_\omega)$. Compared with $\frakF_1(\mD_\omega)$ (see \appref{app:dephasing-upper}), $\frakF_{\textsc{sql}}(\mD_\omega)$ has a factor of $1/(4p(1-p))$ enhancement when we estimate the phase parameter ($\dot p = 0$). When we estimate the noise parameter ($\dot \phi = 0$), however, $\frakF_\sql(\mD_\omega) = \frakF_1(\mD_\omega)$. In general, $\frakF_\sql/\frakF_1$ is between 1 and $1/(4p(1-p))$. 

To sum up, we proved \thmref{thm:HL} and \thmref{thm:SQL} are true for dephasing channels. We also compare our results in this section with the known results in \tableref{table:compare}. The ancilla is not required here. When the noise is non-zero, the QFI must follow the SQL and there exists a spin-squeezed state achieving the QFI asymptotically. In particular, the squeezing parameter should be tuned carefully such that both the $J_x$ and $J_y$ variance are small such that both the noise and the phase parameter are estimated with the optimal precision.

\begin{table}[tb]
\begin{center}
 \begin{tabular}{|| c | c | c ||}  
 \hline
  Case  &  Channel  &  Attainability of \eqref{eq:dephasing-bounds} \\
 \hline\hline
  $H \notin \mS$ ($p = 0$) & $\mD_\omega = \mU_\omega$ &  attainable (see e.g.~\cite{giovannetti2006quantum})\\
 \hline
  $H \in \mS$ ($p \neq 0$) & $\mD_\omega$ when $\dot p = 0$ & attainable (see e.g.~\cite{ulam2001spin}) \\ 
 \hline 
  $H \in \mS$ ($p \neq 0$) & $\mD_\omega$ when $\dot \phi = 0$ & attainable (see e.g.~\cite{fujiwara2003quantum}) \\
 \hline
  $H \in \mS$ ($p \neq 0$)  & general $\mD_\omega$ & unknown until this work\\
 \hline 
\end{tabular}
\end{center}
\caption{Comparison between \secref{sec:dephasing} and previous works}
\label{table:compare}
\end{table}

\section{The QEC protocol}
\label{sec:QEC}

In this section, we introduce a QEC protocol such that every quantum channel simulates the dephasing channel introduced in \secref{sec:dephasing}. To be specific, we find the encoding channel $\mE_{\rm enc}$ and the recovery channel $\mR$ such that 
\begin{equation}
 \mR \circ \mE_\omega \circ \mE_{\rm enc} = \mD_{\tl,\omega}. 
\end{equation}

The construction fully utilizes the advantage of the ancilla, which has the same dimension as the probe with an extra qubit. Let $\dim \mH_\mP = d$ and $\dim \mH_\mA = 2d$. We pick a QEC code 
\begin{equation}
\label{eq:enc}
\ket{0_\tl} \!=\! \sum_{i,j=1}^d \! A_{0,ij} \ket{i}_\mP\!\ket{j,0}_\mA, 
\;
\ket{1_\tl} \!=\! \sum_{i,j=1}^d \! A_{1,ij} \ket{i}_\mP\!\ket{j,1}_\mA,
\end{equation}
with the encoding channel is $\mE_{\rm enc}(\cdot) = V(\cdot)V^\dagger$ where $V = \ket{0_\tl}\bra{0} + \ket{1_\tl}\bra{1}$, and a recovery channel 
\begin{multline}
\label{eq:dec}
\mR(\cdot) = \sum_{m=1}^M \left(\ket{0}\bra{R_m,0} + \ket{1}\bra{Q_m,1}\right) (\cdot) \\ \left(\ket{R_m,0}\bra{0} + \ket{Q_m,1}\bra{1}\right). 
\end{multline}
Here $A_{0,1}$ are matrices in $\bC^{d\times d}$ satisfying $\trace(A_{0,1}^\dagger A_{0,1}) = 1$, $R = (\ket{R_1}\cdots\ket{R_M})$ and $Q = (\ket{Q_1}\cdots\ket{Q_M})$ are matrices satisfying $RR^\dagger = QQ^\dagger = I$. The last ancillary qubit in $\mH_\mA$ guarantees the logical channel to be dephasing, which satisfies
\begin{equation}
\xi = \sum_{i,m} \bra{R_m,0} K_i \ket{0_\tl}\bra{1_\tl} K_i^\dagger \ket{Q_m,1},
\end{equation}
and $\frakF_{\hl,\sql}(\mD_{\tl,\omega})$ could then be directly calculated using \eqref{eq:dephasing-bounds}. Note that in the equation above and in what follows  we use $K_i$ as a substitute for $K_i \otimes I$ for the simplicity of notations.  Below, we will show that by optimizing $\frakF_{\hl,\sql}(\mD_{\tl,\omega})$ over both the recovery channel ($R$,$Q$) and the QEC code ($A_{0,1}$), the QFI upper bounds $\frakF^{(u)}_{\hl,\sql}(\mE_\omega)$ are achievable. 

\section{Achieving the HL upper bound}
\label{sec:HL}

When $H\notin \mS$, we construct a QEC code such that the HL upper bound $\frakF_\hl^{(u)}(\mE_\omega)$ is achieved. For dephasing channels, the HL is achievable only if $\abs{\xi} = 1$. 
Since any transformation $R \leftarrow e^{i\varphi} R$ does not affect the QFI, without loss of generality (WLOG), we assume $\xi = 1$. It means that the QEC has to be perfect, i.e. satisfies the Knill-Laflamme condition~\cite{knill1997theory}
\begin{equation}
P K_i^\dagger K_j P \propto P,\quad \forall i,j,
\end{equation}
where $P = \ket{0_\tl}\bra{0_\tl} + \ket{1_\tl}\bra{1_\tl}$. 
Moreover, there exists a Kraus representation $\{K_i'\}_{i=1}^{r'}$ such that 
$P K'^\dagger_i K'_j P = \mu_i \delta_{ij} P$ and $K'_i P = U_i \sqrt{\mu_i} P$. The unitary $U_i$ has the form
\begin{equation}
U_i = U_{0,i}\otimes\ket{0}\bra{0} + U_{1,i}\otimes\ket{1}\bra{1},
\end{equation}
where $U_{0,i}$ and $U_{1,i}$ are also unitary. 
Let \begin{equation}
\ket{R_i} = \bra{0}U_i\ket{0_\tl}
,
\quad
\ket{Q_i} = \bra{0}U_i\ket{0_\tl},
\end{equation}
for $1 \leq i \leq r'$. We could also add some additional $\ket{R_i}$ and $\ket{Q_i}$ to them to make sure they are two complete and orthonormal bases. Then one could verify that $\xi = 1$ and 
\begin{equation}
\dot{\xi} = -i \trace( (H \otimes I) \sigma_{z,\tl}),
\end{equation}
where $\sigma_{z,\tl} = \ket{0_\tl}\bra{0_\tl} - \ket{1_\tl}\bra{1_\tl}$. Let $\tilde C = A_0 A_0^\dagger - A_1 A_1^\dagger$, $\dot\xi = -i\trace(H\tilde C)$. Then the Knill-Laflamme condition is equivalent to $\trace(\tilde C S) = 0$,  $\forall S\in \mS$. The optimization of the QFI over the QEC code becomes 
\begin{align}
&\text{maximize}~  |\dot\xi| = | \trace(H\tilde C) |,\\
&\text{subject to}~  \|\tilde C\|_1 \leq 2,\,\trace(\tilde C S) = 0,\,\forall \tilde C \in \bH_d,\,S \in \mS,
\end{align}
where $\norm{\cdot}_1$ is the trace norm. 
A similar SDP problem was considered in \cite{zhou2018achieving}. The optimal $|\dot\xi|$ is equal to $2 \min_{S\in \mS} \norm{H-S}$ and the optimal $\tilde C$ could be found via an SDP. Any $A_0,A_1$ such that $\tilde C$ is optimal would achieve the optimal QFI. It means there exists an encoding, and therefore an optimal input state $\ket{\psi_N}$ which is the logical GHZ state, such that 
\begin{equation}
\lim_{N\rightarrow \infty} \frac{F((\mE_\omega^{\otimes N} \otimes \id)(\ket{\psi_{N}}))}{N^2} = 4 \min_{S\in \mS} \norm{H-S}^2.
\end{equation}
Clearly, $4 \min_{S\in \mS} \norm{H-S}^2 = 4 \min_h \norm{\beta}^2 = \frakF^{(u)}_\hl(\mE_\omega)$, {where we used the fact that for any $S \in \mS$ there exists an $h\in \bH_r$ such that $S = \vK^\dagger h \vK$ and vice versa.} \thmref{thm:HL} is then proven. 
Note that, given the optimal $\tilde C$, we can always choose $A_0A_0^\dagger$ and $A_1A_1^\dagger$ with orthogonal supports and the last ancillary qubit in $\mH_\mA$ could be removed because $\ket{0_\tl}$ and $\ket{1_\tl}$ in this case could be distinguished using projections onto the orthogonal supports in $\mH_\mA$~\cite{zhou2018achieving}. Therefore a $d$-dimensional ancillary system is sufficient. 

We have demonstrated the QEC code achieving the optimal HL for arbitrary quantum channels. The code is designed to satisfy the Knill-Laflamme condition and optimize the QFI. 
The logical dephasing channel is exactly the identity channel at the true value of $\omega$ and any change in $\omega$ results in a detectable phase, allowing it to be estimated at the HL. 

\section{Achieving the SQL upper bound}
\label{sec:SQL}

When $H \in \mS$, the situation is much more complicated because when $\abs{\xi} = 1$ we must also have $|\dot\xi| = 0$ and no signal could be detected. Therefore we must consider the trade-off between maximizing the signal and minimizing the noise. To be exact, we want to maximize 
\begin{equation}
\label{eq:sql}
\frakF_\sql(\mD_{\tl,\omega}) = \frac{|\dot\xi|^2}{1 - |\xi|^2}.
\end{equation}
We will show for any $\eta > 0$, there exists a near-optimal code and recovery such that $\frakF_\sql(\mD_{\tl,\omega}) > \frakF_\sql^{(u)}(\mE_{\omega}) - \eta$, proving \thmref{thm:SQL}. We only consider the case where $\frakF_\sql(\mE_\omega) > \frakF_1(\mE_\omega) > 0$ because otherwise $\frakF_1(\mE_\omega) = \frakF_\sql(\mE_\omega)$ and product states are sufficient to achieve $\frakF_\sql(\mE_\omega)$. Detailed derivations could be found in \appref{app:sql} and we sketch the proof here.  To simplify the calculation, we consider a special type of code, the perturbation code, first introduced in \cite{zhou2019optimal}, where 
\begin{equation}
\label{eq:code}
A_{0} = \sqrt{1-\epsilon^2}C + \epsilon D, \quad A_{1} = \sqrt{1-\epsilon^2} C - \epsilon D,
\end{equation}
satisfying $\trace(C^\dagger D) = 0$ and $\trace(C^\dagger C) = \trace(D^\dagger D) = 1$. In this section, we define $\tilde C = C D^\dagger + D C^\dagger$ {(differed by a factor of $\epsilon\sqrt{1-\epsilon^2}$ from the $\tilde C$ defined in \secref{sec:HL})} and also assume $C$ is full rank so that $\tilde C$ could be an arbitrary Hermitian matrix. $\epsilon$ is a small constant and we will calculate $\frakF_\sql(\mD_{\tl,\omega})$ up to the lowest order of $\epsilon$. We adopt the small $\epsilon$ treatment because it allows us to mathematically simplify the optimization of \eqref{eq:sql}, though it is surprising that the optimal QFI is achievable in such a regime where both the signal and the noise are small. Heuristically, it comes from an observation that sometimes the absolute strengths of the signal and the noise are not important---they could cancel each other out in the numerator and the denominator and only the ratio between them matters. See \citep[Appx. G]{zhou2019optimal} for an example. 

To proceed, we first introduce the vectorization of matrices $\dket{\star} = \sum_{ij} \star_{ij} \ket{i}\ket{j}$ for all $\star \in \bC^{d\times d}$ to simplify the notations. We define $E_{0,1}, E, F\in \bC^{d^2 \times r}$ in the following way: 
\begin{equation}
E_{0,1} = (\dket{K_1A_{0,1}} \cdots \dket{K_r A_{0,1}}\big),
\end{equation}
\begin{equation}
\label{eq:ef}
E = (\dket{K_1C} \cdots \dket{K_rC}\big),\,
F = (\dket{K_1D} \cdots \dket{K_rD}\big),
\end{equation}
which satisfy $E_{0,1} = \sqrt{1-\epsilon^2} \pm \epsilon F$, $\trace(E^\dagger F) = 0$ and $\trace(E^\dagger E) = \trace(F^\dagger F) = 1$. Let the recovery matrix $T = Q R^\dagger \in \bC^{d^2 \times d^2}$, then
\begin{equation}
\label{eq:xi}
\xi = \trace(T E_0 E_1^\dagger),\;\;\; \dot\xi = \trace(T \dot E_0 E_1^\dagger) + \trace(T E_0 \dot E_1^\dagger). 
\end{equation}

We consider the regime where both the signal and the noise are sufficiently small---both the denominator and the numerator in \eqref{eq:sql} will be $O(\epsilon^2)$. The recovery matrix $T$ should also be close to the identity operator. We assume $T = e^{i\epsilon G}$ where $G$ is Hermitian and let $\sigma= EE^\dagger$, $\tilde\sigma = i(FE^\dagger - EF^\dagger)$. Expanding $T,E_0,E_1$ around $\epsilon = 0$, we first optimize $\frakF_\sql(\mD_{\tl,\omega})$ over all possible $G$:  
\begin{multline}
\label{eq:sql-G}
\frakF_\sql(\mD_{\tl,\omega})
\approx \\ \max_G \frac{\abs{\trace(G \dot\sigma)}^2 }{4 - 2 \trace(G \tilde \sigma) + \trace(G^2 \sigma) - \abs{\trace(G \sigma)}^2}, 
\end{multline}
up to the lowest order of $\epsilon$. 
The maximization could be calculated by taking the derivative w.r.t. $G$. We can show that the optimal $G$ is 
\begin{equation}
\label{eq:sql-G-opt}
G_{\text{opt}} = \frac{(4  - \trace(L_{\sigma}[\tilde \sigma]\tilde \sigma))}{\trace(L_{\sigma}[\dot \sigma]\tilde \sigma)} L_\sigma[\dot\sigma] + L_\sigma[\tilde\sigma],
\end{equation}
and the corresponding optimal QFI is 
\begin{equation}
\label{eq:sql-sigma}
\frakF_\sql(\mD_{\tl,\omega})
\approx 
\trace(L_\sigma[\dot\sigma]\dot{\sigma}) + \frac{\trace(L_\sigma[\dot\sigma]\tilde{\sigma})^2}{4 - \trace(L_\sigma[\tilde\sigma]\tilde{\sigma})}.
\end{equation}
Now $\frakF_\sql(\mD_{\tl,\omega})$ is a function of the code ($C$ and $D$) only. We will further simplify by writing it as a function of  only $C$ and $\tilde C$. Let $\tau = E^\dagger E$, $\tilde\tau = E^\dagger F + F^\dagger E$, $\tau' = i E^\dagger \dot E -i \dot E^\dagger E$ such that
\begin{gather}
\tau_{ij} = \trace(C^\dagger K_i^\dagger K_j C),\;\tilde\tau_{ij} = \trace(\tilde C K_i^\dagger K_j),\\
\tau'_{ij} = i\trace(C^\dagger K_i^\dagger \dot K_j C) - i\trace(C^\dagger \dot K_i^\dagger K_j C). 
\end{gather}
Then we can verify that 
\begin{align}
\label{eq:tau-1}
\trace(L_\sigma[\dot\sigma]\dot{\sigma}) &= 4 \trace(C^\dagger \dot\vK^\dagger \dot\vK C) - \trace(L_\tau[\tau']\tau'),\\
\trace(L_\sigma[\dot\sigma]\tilde{\sigma}) &= - 2 \trace(\tilde C H) + \trace(L_\tau[\tau']\tilde\tau),\\
\label{eq:tau-3}
\trace(L_\sigma[\tilde\sigma]\tilde{\sigma}) &= 4 - \trace(L_\tau[\tilde\tau]\tilde\tau).
\end{align}
and 
\begin{equation}
\begin{split}
&\frakF_\sql(\mD_{\tl,\omega}) \approx f(C,\tilde C) = 4 \trace(C^\dagger \dot\vK^\dagger \dot\vK C) \\
&\quad - \trace(L_\tau[\tau']\tau')  + \frac{(- 2 \trace(\tilde C H)  + \trace(L_\tau[\tau']\tilde\tau))^2}{\trace(L_\tau[\tilde\tau]\tilde\tau)}. 
\end{split}
\end{equation}

At this stage, it is not obvious why the maximization of $\frakF_\sql(\mD_{\tl,\omega})$ over $C$ and $\tilde C$ is equal to $\frakF^{(u)}_\sql(\mE_\omega)$. To see that, we need to reformulate the SQL upper bound using its dual program. First we note that 
\begin{equation}
\frakF^{(u)}_\sql(\mE_\omega) = \max_{C:\trace(C^\dagger C) = 1}\min_{h:\beta = 0}4\trace(C^\dagger \alpha C),
\end{equation}
where we are allowed to exchange the order of maximization and minimization thanks to Sion's minimax theorem~\cite{komiya1988elementary,do2001introduction}. Fixing $C$, we consider the optimization problem $\min_{h:\beta=0} 4\trace(C^\dagger \alpha C)$. When $C$ is full rank, we can show that it is equivalent to $\max_{\tilde C\in \bH_d} f(C,\tilde C)$, 
where $\tilde C$ is introduced as the Lagrange multiplier associated with the constraint $\beta = 0$~\cite{boyd2004convex} and the optimal $\tilde C$ is traceless. 

The procedure to find a near-optimal code such that $\frakF_\sql(\mD_{\tl,\omega}) > \frakF^{(u)}_\sql(\mE_\omega) - \eta$ for any $\eta > 0$ goes as follows:
\begin{enumerate}[(1),wide,labelwidth=!,labelindent=0pt]
\item Find a full rank $C^\diamond$ such that $\trace(C^{\diamond\dagger}C^\diamond) = 1$ and 
$
\min_{h:\beta = 0}4\trace(C^{\diamond\dagger} \alpha C^\diamond) > \frakF^{(u)}_\sql(\mE_\omega) - \eta/2.
$
\item Find a Hermitian $\tilde C^\diamond$ such that $f(C^\diamond,\tilde C)$ is maximized 
and let $D^\diamond = \frac{1}{2}C^{\diamond-1}\tilde C^\diamond$. Rescale $D^\diamond$ such that $\trace(D^{\diamond\dagger}D^\diamond) = 1$.
\item Calculate $\frakF_\sql(\mD_{\tl,\omega})|_{C=C^\diamond,D=D^\diamond}$ using \eqsref{eq:code}{eq:xi} and \eqref{eq:sql-G-opt}. Find a small $\epsilon^\diamond > 0$ such that 
$
\frakF_\sql(\mD_{\tl,\omega}) > f(C^\diamond,\tilde C^\diamond) - \eta/2.
$
\end{enumerate}
The numerical algorithms for step (1) and (2) are provided in \appref{app:algo}, where the most computationally intensive part is an SDP. Note that in contrast to the HL case, here we require $2d$-dimensional ancillas, twice as large as probes. In principle, however, $d$-dimensional ancillas are sufficient to achieve the asymptotic QFI, considering the Schmidt decomposition on the input state, though we no longer have explicit encoding and decoding protocols when using $d$-dimensional ancillas.

To conclude, we proposed a perturbation code which could achieve the SQL upper bound with an arbitrarily small error. We take the limit where the parameter $\epsilon$ which distinguishes the logical zero and one states is sufficiently small. Note that if we take $\epsilon = 0$, the probe state will be a product state and we can only achieve $\frakF_1(\mD_{\tl,\omega})$. This discontinuity appears because we must first take the limit $N \rightarrow \infty$ before taking the limit $\epsilon \rightarrow 0$ and the impact of a small $\epsilon$ becomes significant in the asymptotic limit.

\section{Examples}


In this section, we provide three applications of our theorems. We first compute the asymptotic QFI of single-qubit depolarizing channels, which were not fully explored before. It is a case where $\frakF_\sql > \frakF_1$ whenever the HNKS condition is violated. Secondly, we consider amplitude damping channels and obtain an analytical solution of a near-optimal QEC protocol. We will directly see how the gap between the attainable QFI and $\frakF_\sql$ shrinks when $\epsilon$ approaches $0$. In the third example, we consider a special type of channel which always satisfies $\frakF_\sql = \frakF_1$ and provide a new simple proof of it.  

\subsection{Single-qubit depolarizing channels}
\label{sec:depolarizing}

Here we calculate $\frakF_1$, $\frakF_\sql$ and $\frakF_\hl$ for depolarizing channels $\mN^{{\rm d}}_{\omega}(\rho) = \mN^{{\rm d}}(\mU_\omega(\rho))$ where 
\begin{equation}
\mN^{{\rm d}}(\rho) = (1-p) \rho  +  p_x \sigma_x \rho \sigma_x + p_y \sigma_y \rho \sigma_y + p_z \sigma_z \rho \sigma_z,
\end{equation}
$p_{x,y,z} \geq 0$, $p = p_x + p_y + p_z < 1$ and $\mU_\omega(\cdot) = e^{-\frac{i\omega}{2}\sigma_z}(\cdot)e^{\frac{i\omega}{2}\sigma_z}$. $p_{x,y,z}$ are independent of $\omega$. 

First, we notice that HNKS is satisfied if and only if $p_x = p_z = 0$ or $p_y = p_z = 0$. When HNKS is satisfied, $\frakF_\hl(\mN^{{\rm d}}_\omega) = 1$. It is the same as the $\frakF_\hl$ when there is no noise ($p=0$) because the Kraus operator ($\sigma_x$ or $\sigma_y$) is perpendicular to the Hamiltonian ($\sigma_z$) and could be fully corrected. It is consistent with previous results for single-qubit Hamiltonian estimation that the HL is achievable if and only if the Markovian noise is rank-one and not parallel to the Hamiltonian~\cite{kessler2014quantum,arrad2014increasing,dur2014improved,ozeri2013heisenberg,unden2016quantum,reiter2017dissipative,sekatski2017quantum}. As calculated in \appref{app:depolarizing}, 
\begin{equation}
\frakF_1(\mN^{{\rm d}}_\omega) = 1 - w,
\end{equation}
where $w = 4 \left(\frac{p_xp_y}{p_x+p_y} + \frac{(1-p)p_z}{1-p+p_z}\right) \leq 1$. 
When HNKS is violated, 
\begin{equation}
\frakF_\sql(\mN^{{\rm d}}_\omega) = (1 - w)/w.
\end{equation}
In the equations above, when $p_x = p_y = 0$, we take $\frac{p_xp_y}{p_x+p_y} = 0$, in which case $\mN^{{\rm d}}_\omega$ becomes the dephasing channel introduced in \secref{sec:dephasing} where $\phi = \omega$ and $p$ is independent of $\omega$. 

We observe that 
\begin{equation}
\frakF_\sql(\mN^{{\rm d}}_\omega) = \frakF_1(\mN^{{\rm d}}_\omega)/w \geq \frakF_1(\mN^{{\rm d}}_\omega),
\end{equation}
and the equality ($w = 1$) holds if and only if $p_x = p_y$ and $p_z + p_x = 1/2$, in which case $\frakF_\sql(\mN^{{\rm d}}_\omega) = \frakF_1(\mN^{{\rm d}}_\omega) = 0$ and $\mN^{{\rm d}}_\omega = \mN^{{\rm d}}$ becomes a mixture of a completely dephasing channel and a completely depolarizing channel~\cite{watrous2018theory} where $\omega$ cannot be detected.

\begin{figure}[tb]
\includegraphics[width=0.4\textwidth]{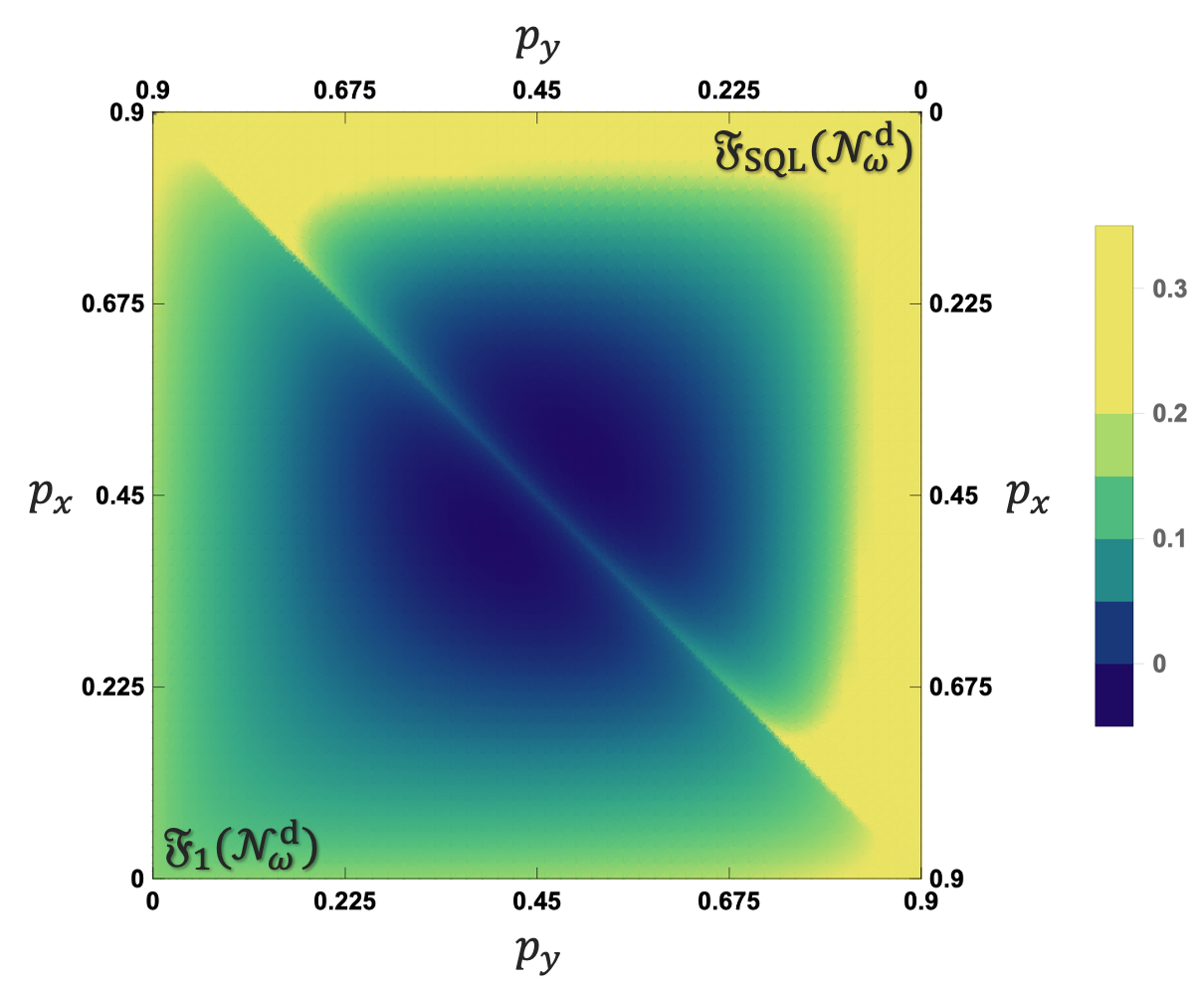}
\caption{\label{fig:depolarizing} 
Plots of $\frakF_1(\mN^{{\rm d}}_\omega)$ and $\frakF_\sql(\mN^{{\rm d}}_\omega)$ as functions of $p_x$ and $p_y$ when $p_z = 0.1$. The lower left and upper right part are the plots of $\frakF_1(\mN^{{\rm d}}_\omega)$ and $\frakF_\sql(\mN^{{\rm d}}_\omega)$ respectively. 
}
\end{figure}

$\frakF_\sql(\mN^{{\rm d}}_\omega)$ is in general non-additive. In particular, when $p \ll 1$, we have $w \ll 1$ 
and $\frakF_\sql(\mN^{{\rm d}}_\omega) \gg \frakF_1(\mN^{{\rm d}}_\omega)$. We also illustrate the difference between $\frakF_\sql(\mN^{{\rm d}}_\omega)$ and $\frakF_1(\mN^{{\rm d}}_\omega)$ in \figref{fig:depolarizing} by plotting $\frakF_\sql(\mN^{{\rm d}}_\omega)$, $\frakF_1(\mN^{{\rm d}}_\omega)$ as a function of $p_x$ and $p_y$ when $p_z = 0.1$. $\frakF_\sql(\mN^{{\rm d}}_\omega) = \frakF_1(\mN^{{\rm d}}_\omega) = 0$ at $(p_x,p_y,p_z) = (0.4,0.4,0.1)$. The ratio $\frakF_\sql(\mN^{{\rm d}}_\omega)/\frakF_1(\mN^{{\rm d}}_\omega)$ increases near the boundary of $p_x + p_y < 0.9$. 

We remark here that when the dimension of the system is large, for example, a qudit depolarizing channel or a collective dephasing channel~\cite{dorner2012quantum}, although $\frakF_1$, $\frakF_\sql$ and the optimal input states can be found numerically via SDPs, analytical solutions may not exist. In that case, it might be helpful to compute analytical upper bounds on the QFI~\citep[Appx. F]{zhou2020new} or consider the limit of large ensembles and use variational methods to solve for the QFI~\cite{knysh2014true}.

\subsection{Amplitude damping channels}

In the first example, we focus on computing the asymptotic QFIs for single-qubit depolarizing channels, but we do not provide explicit QEC protocols achieving the QFIs. Here we present a second example, where we obtain an analytical solution of the optimal QEC protocol and also analyze its performance when $\epsilon$ is not a small constant. 

Here we consider amplitude damping channels $\mN^{{\rm ad}}_{\omega}(\rho) = \mN^{{\rm ad}}(\mU_\omega(\rho))$ defined by 
\begin{equation}
\mN^{{\rm ad}}(\rho) = K_1^{{\rm ad}} \rho K_1^{{\rm ad}\dagger} + K_2^{{\rm ad}} \rho K_2^{{\rm ad}\dagger},
\end{equation}
where $K_1^{{\rm ad}} = \ket{0}\bra{0} + \sqrt{1-p}\ket{1}\bra{1}$ and $K_2^{{\rm ad}} = \sqrt{p}\ket{0}\bra{1}$ and $p$ represents the probability of a particle switching from $\ket{1}$ to $\ket{0}$ which is independent of $\omega$. $\mU_\omega$ again is the Pauli-Z rotation $e^{-i\frac{\omega}{2}\sigma_z}$. We will assume $\omega = 0$ in this section for simplicity, because for non-zero $\omega$, the QFI is the same and we only need to rotate the code accordingly. 

As before, amplitude damping channels follow the SQL as long as $p > 0$. Thus, we shall only focus on the situation where HNKS is violated. As shown in \appref{app:ad}, $\frakF_\sql(\mN^{{\rm ad}}_\omega) = 4({1-p})/{p}$~\cite{demkowicz2012elusive} and the near-optimal QEC protocol can be obtained using our algorithm from \secref{sec:SQL}. 
The QEC code is characterized by two small but non-zero constants $\delta$ and $\epsilon$, where $\delta$ is to make sure $C^\diamond$ is full rank, originated from step (1) in our algorithm from \secref{sec:SQL} and $\epsilon = o(\delta)$ is the small constant in the perturbation code:
\begin{gather}
\ket{0_\tl} 
= \sin(\delta + \epsilon)\ket{0}_\mP \!\ket{00}_\mA \!+\! \cos(\delta + \epsilon)\ket{1}_\mP \!\ket{10}_\mA,\\
\ket{1_\tl}  
= \sin(\delta - \epsilon)\ket{0}_\mP \ket{01}_\mA \!+\! \cos(\delta - \epsilon)\ket{1}_\mP \!\ket{11}_\mA. 
\end{gather}
Note that we use trigonometric functions instead of $\epsilon$ and $\sqrt{1-\epsilon^2}$ as before just to simplify the notations. We also need the optimal recovery channel which is determined by 
\begin{equation}
G_{\text{opt}} = \frac{2i}{\sqrt{1-p}}\ket{00}\bra{11} + \frac{-2i}{\sqrt{1-p}}\ket{11}\bra{00}. 
\end{equation}

The asymptotic channel QFI $\frakF_\sql(\mD_{\tl,\omega})$ attainable using the QEC protocol above is $\frakF_\sql(\mD_{\tl,\omega}) = |\dot\xi|^2/(1 - |\xi|^2)$, where 
\begin{gather}
\xi = 1 - \frac{2p\sin^2(\delta)}{1-p} \epsilon^2 + O(\epsilon^4),\\ 
\dot\xi = -2i\sin(2\delta)\epsilon + O(\epsilon^3), 
\end{gather}
and 
\begin{equation}
\label{eq:epsilon-zero}
\frakF_\sql(\mD_{\tl,\omega}) = \frac{4(1-p)\cos^2(\delta)}{p} + O(\epsilon^2). 
\end{equation}
which approaches $\frakF_\sql(\mN^{{\rm ad}}_\omega)$ for small $\delta$. Note, however, that we cannot take $\delta = 0$ because then $\dot\xi = 0$. It means $\frakF_\sql(\mN^{{\rm ad}}_\omega)$ is achievable with an arbitrarily small but non-zero error. The exact values of $\xi$ and $\dot\xi$ as a function of $\delta$ and $\epsilon$ can be found in~\appref{app:ad}.

\begin{figure}[tb]
\includegraphics[width=0.36\textwidth]{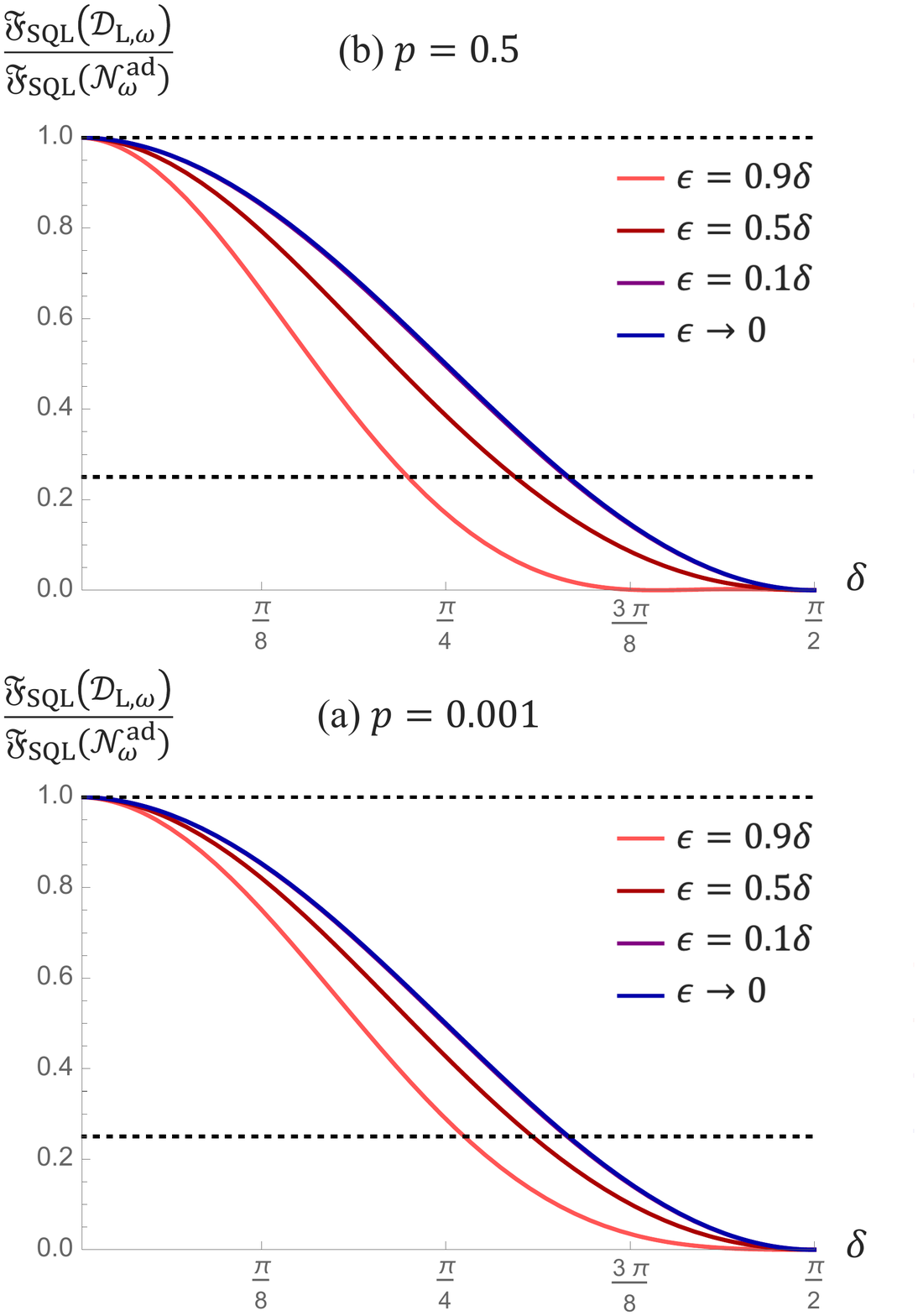}
\caption{\label{fig:ad} 
Plots of $\frakF_\sql(\mD_{\tl,\omega})/\frakF_\sql(\mN^{{\rm ad}}_\omega)$ as a function of $\delta$. We take $p = 0.5, 0.001$ in (a) and (b) and $\epsilon=0.9\delta, 0.5\delta, 0.1\delta$, $\epsilon\rightarrow 0$ in each figure. The curves from $\epsilon = 0.1\delta$ and $\epsilon \rightarrow 0$ are almost indistinguishable from each other. The dashed lines are at $1$ and $1/4$ where the former represents the upper bound $\frakF_\sql(\mN^{{\rm ad}}_\omega)$ and the latter represents the optimal asymptotic QFI without the assistance of ancillas $\frakF_\sql(\mN^{{\rm ad}}_\omega)/4$~\cite{knysh2014true,demkowicz2014using}. Our QEC protocol outperforms the ancilla-free protocols even for large $\delta$ and $\epsilon$. 
}
\end{figure}

To visualize the gap between $\frakF_\sql(\mD_{\tl,\omega})$ and $\frakF_\sql(\mN^{{\rm ad}}_\omega)$, we plot it in \figref{fig:ad}. We take $p = 0.5, 0.001$ in \figaref{fig:ad} and \figbref{fig:ad}, and $\epsilon=0.9\delta,0.5\delta, 0.1\delta$, $\epsilon\rightarrow 0$ in each figure, and plot the ratio between the attainable QFI $\frakF_\sql(\mD_{\tl,\omega})$ and the optimal QFI $\frakF_\sql(\mN^{{\rm ad}}_\omega)$ as a function of $\delta$. \figaref{fig:ad} and \figbref{fig:ad} are almost identical, showing the robustness of our code against the change in noise rates. We also see that the curve from $\epsilon = 0.1\delta$ almost overlaps with the limiting one (\eqref{eq:epsilon-zero}) as $\epsilon \rightarrow 0$. Moreover, we compare our ancilla-assisted QEC protocol with ancilla-free ones which achieve at most $\frakF_\sql(\mN^{{\rm ad}}_\omega)/4$~\cite{knysh2014true,demkowicz2014using}. It outperforms the optimal ancilla-free ones in a large range (of $\delta$ and $\epsilon$), showing the power of noiseless ancillas in phase estimation under amplitude damping noise. This type of phenomenon does not occur in dephasing channels where ancilla-free protocols are optimal~\cite{ulam2001spin,knysh2014true,demkowicz2014using}.

\subsection{\texorpdfstring{$\bU$}{U}-covariant channels}

Finally, we consider a special class of channel which satisfies $\frakF_\sql = \frakF_1$. It covers many typical noise channels such as Pauli channels where noise rates are unknown, whereas in the first two examples we focus on phase estimation. Let $\bU = \{U_i\}_{i=1}^n \subset \bC^{d\times d}$ be a set of unitary operators such that for some probability distribution $\{p_i\}_{i=1}^n$, $\{(p_i,U_i)\}_{i=1}^n$ is a unitary 1-design~\cite{dankert2005efficient}, satisfying
\begin{equation}
\sum_{i=1}^n p_i U_i A U_i^\dagger = \trace(A) \frac{I}{d},\;\forall A \in \bC^{d\times d}.
\end{equation}
For example, when $\bU$ is a unitary orthonormal basis of $\bC^{d\times d}$, $\{(\frac{1}{d^2},U_i)\}_{i=1}^{d^2}$ is a unitary 1-design. 
Given a quantum channel $\mT_\omega(\cdot) = \sum_{i=1}^r K_i (\cdot) K_i^\dagger$, we call it $\bU$-covariant if for all $U \in \bU$, there is a unitary $V$ such that 
\begin{equation}
\label{eq:tele}
\mT_\omega(U\rho U^\dagger) = V \mT_\omega (\rho) V^\dagger. 
\end{equation}
Note that here it is important that $U$ and $V$ are independent of $\omega$, a feature called joint covariance~\cite{laurenza2018channel}. It could be shown that $\frakF_1(\mT_\omega) = \frakF_\sql(\mT_\omega)$ when $\mT_\omega$ is $\bU$-covariant, using the teleportation simulation technique~\cite{pirandola2017ultimate,pirandola2017fundamental,chiribella2009realization,wilde2017converse}. Here we provide an alternative proof using only the definitions of $\frakF_1$ and $\frakF_\sql$ in the minimax formulation.

Let $h^\bdia$ be a solution of $\min_h \max_\rho 4\trace(\rho\alpha)$. As explained in \appref{app:algo}, for every $\rho^\bdia$ which is a solution of $\max_\rho \min_h 4\trace(\rho\alpha)$, $(h^\bdia,\rho^\bdia)$ is a saddle point, i.e.
\begin{equation}
\label{eq:saddle}
 4\trace(\rho\alpha^\bdia) \leq 4\trace(\rho^\bdia\alpha^\bdia) \leq 4\trace(\rho^\bdia \alpha),
\end{equation}
for all $\rho$ and $h$, where $\alpha^\bdia = \alpha|_{h=h^\bdia}$. Then 
$\dket{C^\bdia} \in \mH_\mP \otimes \mH_\mA$ is an optimal input state of a single quantum channel $\mT_\omega$, if and only if $\rho^\bdia = C^\bdia C^{\bdia\dagger}$ satisfies \eqref{eq:saddle}. 
According to \eqref{eq:tele}, if $\dket{C^\bdia}$ is an optimal input, $\dket{U C^\bdia} = (U\otimes I)\dket{C^\bdia}$ is also an optimal input for all $U \in \bU$ and satisfies \eqref{eq:saddle}. Then 
$\sum_{i=1}^n p_i U_i\rho^\bdia U_i^\dagger = \frac{I}{d}$ also satisfies \eqref{eq:saddle}, implying the maximally entangled state $\dket{\frac{I}{d}}$ is an optimal input for $\mT_\omega$. 
The discussion above also works for $\mT_\omega^{\otimes N}$ because $\mT_\omega^{\otimes N}$ is $\bU^{\otimes N}$-covariant and $\{(\Pi_{k}p_{i_k},\otimes_{k} U_{i_k})\}$ is a unitary 1-design on $\bC^{Nd\times Nd}$. Therefore $\dket{\frac{I}{d^N}}$ is an optimal input for $\mT_\omega^{\otimes N}$, which implies $\frakF_N(\mT_\omega) = N \frakF_1(\mT_\omega)$.

\section{Conclusions and outlook}
\label{sec:conclusion}

In this paper, we focus on the asymptotic behaviour of the QFI of a quantum channel when the number of identical channels $N$ is infinitely large. {We consolidate the HNKS condition by showing it unambiguously determines} whether or not the scaling of the asymptotic QFI is quadratic or linear. In both cases, we show that the optimal input state achieving the asymptotic QFI could be found via an SDP. To find the optimal input state, we reduce every quantum channel to a single-qubit dephasing channel where both the phase and the noise parameter vary w.r.t. the unknown parameter and then optimize the asymptotic QFI of the logical dephasing channel over the encoding and the recovery channel. The optimal input state is either the logical GHZ state (when HNKS is satisfied) or the logical spin-squeezed state (when HNKS is violated). This provides a unified framework for channel estimation while previous results were centered on either Hamiltonian estimation or noise estimation in special situations. 

Furthermore, our results implies that when HNKS is violated, sequential strategies provide no advantage over parallel strategies asymptotically, answering another open problem in quantum metrology. The regularized channel QFI $\frakF_\sql(\mE_\omega)$ is a useful information-theoretic measure and was recently shown to be useful in deriving bounds in covariant QEC~\cite{kubica2020using,zhou2020new}. It can also serve as a useful benchmark for practical quantum metrological tasks---one could compare the attainable Fisher information with $\frakF_\sql(\mE_\omega)$ to determine how far a metrological protocol is from optimal. Moreover, we propose a two-dimensional QEC protocol to achieve $\frakF_\sql(\mE_\omega)$, where the optimal input state is a concatenation of many-body spin-squeezed states and two-dimensional QEC codes (\figref{fig:sketch}). It allows us to reduce the optimization in the entire Hilbert space which is exponentially large to that 
in a local Hilbert space, providing a new inspiration for numerical methods in quantum metrology~\cite{gorecki2019quantum,chabuda2020tensor,kaubruegger2019variational,koczor2020variational,meyer2020variational}.

It is left open whether sequential strategies provide no advantage over parallel strategies asymptotically when HNKS is satisfied. The statement was proven true only for unitary channels~\cite{giovannetti2006quantum}, but there is still a gap between $\frakF_\hl(\mE_\omega)$ and the state-of-the-art upper bounds on $\frakF^{(\rm seq)}_\hl(\mE_\omega)$ for general quantum channels~\cite{demkowicz2014using,sekatski2017quantum,yuan2017fidelity,katariya2020geometric}. Note that for multi-parameter estimation, a gap between parallel strategies and sequential strategies exists even for unitary channels~\cite{yuan2016sequential}. 

\section*{Acknowledgments}

We thank Chang-Ling Zou, Mark Wilde, Haining Pan, Zi-Wen Liu, Yuxiang Yang, Rafal Demkowicz-Dobrzanski, Vishal Katariya and Yu Chen for helpful discussions. We acknowledge support from the ARL-CDQI (W911NF-15-2-0067), ARO (W911NF-18-1-0020, W911NF-18-1-0212), ARO MURI (W911NF-16-1-0349), AFOSR MURI (FA9550-15-1-0015, FA9550-19-1-0399), DOE (DE-SC0019406), NSF (EFMA-1640959, OMA-1936118), and the Packard Foundation (2013-39273).

\bibliography{refs-entanglement}{}

\onecolumngrid
\newpage
\appendix

\section{Deriving the upper bound on \texorpdfstring{$\frakF_N(\mE_\omega)$}{F\_N(Epsilon\_omega)}}
\label{app:upper}

For completeness, we provide a proof~\cite{fujiwara2008fibre} of \eqref{eq:upper} in the main text. 
Let $K_i^{(1)} = K_i$ for $i \in [r]$, where $[r] = \{1,2,\ldots,r\}$. Inductively, let  
\begin{equation}
K_{\iota}^{(n+1)} = K_{\iota_1}^{(n)} \otimes K^{(1)}_{\iota_2},\quad 
\forall \iota = (\iota_1,\iota_2) \in [r]^n \times [r]. 
\end{equation}
$\{K_{\iota}^{(n)}\}_{\iota \in [r]^n}$ is a Kraus representation of $\mE_\omega^{\otimes n}$ for all $n$. Then let 
$
\alpha^{(n)} = \sum_{\iota_1}  \dot{K}_{\iota_1}^{(n)\dagger } \dot{K}_{\iota_1}^{(n)},
\beta^{(n)} = i \sum_{\iota_1} K_{\iota_1}^{(n)\dagger } \dot{K}_{\iota_1}^{(n)}
$, 
we have 
\begin{gather}
\alpha^{(n+1)} = \sum_{\iota_1,\iota_2} \bigg( \frac{\partial (K_{\iota_1}^{(n)} \otimes K^{(1)}_{\iota_2}) }{\partial \omega}\bigg)^\dagger \bigg(
\frac{\partial (K_{\iota_1}^{(n)} \otimes K^{(1)}_{\iota_2}) }{\partial \omega} \bigg) = \alpha^{(n)} \otimes I + 2 \beta^{(n)}\otimes \beta^{(1)} + I \otimes \alpha^{(1)},\\
\beta^{(n+1)} = i \sum_{\iota_1,\iota_2} \bigg( \frac{\partial (K_{\iota_1}^{(n)} \otimes K^{(1)}_{\iota_2}) }{\partial \omega}\bigg)^\dagger (K_{\iota_1}^{(n)} \otimes K^{(1)}_{\iota_2}) = \beta^{(n)}\otimes I + I \otimes \beta^{(1)}. 
\end{gather}
The solution is $\beta^{(N)} = \sum_{k=0}^{N-1} I^{\otimes k}\otimes \beta^{(1)} \otimes I^{\otimes N-1-k}$ and 
\begin{equation}
\alpha^{(N)} = \sum_{k=0}^{N-1} I^{\otimes k}\otimes \alpha^{(1)} \otimes I^{\otimes N-1-k} + 2 \sum_{k_1=0}^{N-2}\sum_{k_2=0}^{N-2-k_1} I^{\otimes k_1} \otimes \beta^{(1)} \otimes I^{\otimes k_2} \otimes \beta^{(1)} \otimes  I^{\otimes N-2-k_1-k_2} . 
\end{equation}
Therefore, $\frakF_N(\mE_\omega) \leq 4 \|\alpha^{(N)}\| \leq 4 N \|\alpha^{(1)}\| + 4 N(N-1) \|\beta^{(1)}\|^2$
 and the inequality holds for any Kraus representation of $\mE_\omega$. We can choose $\vK' = u \vK$, then 
\begin{equation}
\frakF_N(\mE_\omega) \leq 4 \min_h \left( N \|\alpha\| + N(N-1) \|\beta\|^2 \right), 
\end{equation}
where $h = i u^\dagger \dot u$ is an arbitrary Hermitian matrix, $\alpha = \dot\vK'^\dagger \dot\vK' = (\dot\vK - ih\vK)^\dagger (\dot\vK-ih\vK)$ and  $\beta = i \vK'^\dagger \dot\vK' = i \vK^\dagger (\dot\vK-ih\vK)$ .

\section{Calculating the QFI upper bounds for dephasing channels}
\label{app:dephasing-upper}

Here we calculate $\frakF_\hl^{(u)} = 4 \min_{h}\norm{\beta}^2$ and $\frakF_\sql^{(u)} = 4 \min_{h:\beta = 0}\norm{\alpha}$ for dephasing channels
\begin{equation}
\mD_\omega(\rho) = (1 - p) e^{- \frac{i \phi}{2} \sigma_z} \rho e^{\frac{i \phi}{2} \sigma_z} + p \sigma_z e^{- \frac{i \phi}{2} \sigma_z} \rho e^{\frac{i \phi}{2} \sigma_z} \sigma_z = \sum_{i=1}^2 K_i \rho K_i^\dagger.
\end{equation}
where $K_1 = \sqrt{1-p} e^{- \frac{i \phi}{2} \sigma_z}, K_2 = \sqrt{p} \sigma_z e^{- \frac{i \phi}{2} \sigma_z}$. Assume $p > 0$, then 
\begin{gather}
\vK = 
\begin{pmatrix}
\sqrt{1-p} e^{- \frac{i \phi}{2} \sigma_z}\\
\sqrt{p} \sigma_z e^{- \frac{i \phi}{2} \sigma_z}
\end{pmatrix}
,\quad 
\dot\vK = 
\begin{pmatrix}
\big( \frac{-\dot p}{2\sqrt{1-p}} - \sqrt{1-p} \frac{i\dot\phi}{2} \sigma_z \big) e^{- \frac{i \phi}{2} \sigma_z}\\
\big( \frac{\dot p}{2\sqrt{p}} - \sqrt{p} \frac{i\dot\phi}{2} \sigma_z \big) e^{- \frac{i \phi}{2} \sigma_z}\sigma_z
\end{pmatrix}
,\\
\dot\vK - i h \vK = 
\begin{pmatrix}
\big( \frac{-\dot p}{2\sqrt{1-p}} -ih_{11}\sqrt{1-p} - \sqrt{1-p} \frac{i\dot\phi}{2} \sigma_z - ih_{12}\sqrt{p}\sigma_z\big) e^{- \frac{i \phi}{2} \sigma_z}\\
\big( \frac{\dot p}{2\sqrt{p}}  \sigma_z - ih_{22} \sqrt{p} \sigma_z - \sqrt{p} \frac{i\dot\phi}{2} - i h_{21} \sqrt{1-p}\big) e^{- \frac{i \phi}{2} \sigma_z}
\end{pmatrix},\\
\beta = i\vK^\dagger (\dot\vK - ih\vK) = \frac{\dot\phi}{2}\sigma_z + (1-p)h_{11} + p h_{22} + \sqrt{p(1-p)}(h_{12}+h_{21})\sigma_z,\\
\begin{split}
\alpha 
&= (\dot\vK - ih\vK)^\dagger (\dot\vK-ih\vK) \\
&= \frac{\dot p^2}{4p(1-p)} + h_{11}^2 (1-p) + h_{22}^2 p + \frac{\dot\phi^2}{4} + \abs{h_{12}}^2 + 2\sqrt{p(1-p)}\dot\phi \Re[h_{12}] \\
&\quad + 2\Re\Big[ -\frac{\dot p\sqrt{p}}{\sqrt{1-p}} i h_{12} + ((1-p)h_{11} + h_{22}p)\frac{\dot\phi}{2} + (h_{11}h_{12} + h_{22}h_{21})\sqrt{p(1-p)} - i \frac{\dot p\sqrt{1-p}}{\sqrt{p}} h_{21}\Big] \sigma_z.
\end{split}
\end{gather}

$\beta = 0$ is equivalent to 
$(1-p)h_{11} + p h_{22} = 0$ and $\frac{\dot\phi}{2} + \sqrt{p(1-p)} (h_{12}+h_{21}) = 0$, which is achievable for any $p > 0$. 
When $h_{11}=h_{22} =0$ and $h_{12} = h_{21} = - \frac{\dot\phi}{4\sqrt{p(1-p)}}$, $\norm{\alpha} = \min_{h:\beta = 0} \norm{\alpha} = \frac{(1-2p)^2\dot\phi^2}{16p(1-p)} + \frac{\dot p^2}{4(1-p)p}$. Then
\begin{equation}
\frakF_\sql^{(u)}(\mD_\omega) = 4 \min_{h:\beta = 0} \norm{\alpha} = \frac{(1-2p)^2\dot\phi^2}{4p(1-p)} + \frac{\dot p^2}{(1-p)p} = \frac{|\dot\xi|^2}{1-\abs{\xi}^2},
\end{equation}
where $\xi = (1-2p)e^{-i\phi} = \bra{0}\mD_\omega(\ket{0}\bra{1})\ket{1}$ is a complex number completely determining the channel. 

When $p = 0$, we must also have $\dot p =0$. Then $\beta = \frac{\dot\phi}{2} \sigma_z + h_{11}$ and 
\begin{equation}
\frakF_\hl^{(u)}(\mD_\omega) = 4 \min_{h} \norm{\beta}^2 = |\dot\phi|^2 = |\dot\xi|^2. 
\end{equation}

We can also calculate the channel QFI
\begin{equation}
\frakF_1^{(u)}(\mD_\omega) = 4 \min_{h} \norm{\alpha} = 
\begin{cases}
{(1-2p)^2} \dot\phi^2 + \frac{\dot p^2}{(1-p)p}, & p > 0,\\
{(1-2p)^2} \dot\phi^2, & p = 0.
\end{cases}
\end{equation}
It could be achieved using $\ket{\psi_0} = \frac{\ket{0}+\ket{1}}{\sqrt{2}}$. 

\section{A useful formula for calculating the QFI of dephasing channels}
\label{app:dephasing-sum}

In this appendix, we prove \eqref{eq:dephasing-sum} in the main text. 
Let $\ket{\psi} = e^{- i\phi J_z} \ket{\psi_0}$ and a subspace 
\begin{equation}
\mZ = {\rm span}\Big\{\prod_{k=1}^N (\sigma_z^{(k)})^{j_k}\ket{\psi} ,(j_1,\ldots,j_N)\in\{0,1\}^{N}\Big\}.
\end{equation}
Assume $\dim \mZ = n$. $\mZ$ must have an orthonormal basis $\{\ket{e_\ell}\}_{\ell=1}^n$ where $\ket{e_\ell} = \sum_{j_1,\ldots,j_N=0}^1 r_{\ell,(j_1,\ldots,j_N)} \prod_{k=1}^N (\sigma_z^{(k)})^{j_k} \ket{\psi}$ with real $r_{\ell,(j_1,\ldots,j_N)}$. For example, one can use the Gram-Schmidt procedure to find $\{\ket{e_\ell}\}_{\ell=1}^n$ because $\bra{\psi}\prod_{k=1}^N (\sigma_z^{(k)})^{j_k}\ket{\psi} \in \mathbb{R}$ for all $(j_1,\ldots,j_N)\in\{0,1\}^{\otimes N}$. 

Then
\begin{equation}
\begin{split}
\rho_\omega 
&= \mD_\omega^{\otimes N} (\ket{\psi_0}\bra{\psi_0}) = (\mD_\omega|_{\phi = 0})^{\otimes N} (\ket{\psi}\bra{\psi}) \\
&= \sum_{j_1,\ldots,j_N=0}^1 (1-p)^{(N - \sum_{k=1}^N j_k)}p^{(\sum_{k=1}^N j_k)}\prod_{k=1}^N (\sigma_z^{(k)})^{j_k} \ket{\psi} \bra{\psi} \prod_{k=1}^N (\sigma_z^{(k)})^{j_k} 
= \sum_{\ell,\ell'=1}^n \chi_{\ell\ell'} \ket{e_\ell}\bra{e_{\ell'}} 
\end{split}
\end{equation}
where $\chi \in \bR^{n \times n}$ is a symmetric matrix. $\chi = \sum_{i=1}^n \mu_{i} v_i v_i^T$ where $v_i$ are real orthonormal eigenvectors of $\chi$. Then we can write $\rho_\omega = \sum_{\ell=1}^n \mu_\ell \ket{\psi_\ell}\bra{\psi_{\ell}}$ where $\ket{\psi_{\ell}} = \sum_{\ell'=1}^n v_{\ell\ell'}\ket{e_{\ell'}}$.
Then according to the definition of QFI, 
\begin{equation}
\label{eq:dephasing-QFI}
F(\rho_\omega) = 2 \sum_{\ell \ell':\mu_\ell + \mu_{\ell'} \neq 0} \frac{\abs{\bra{\psi_{\ell}}\dot{\rho}_\omega\ket{\psi_{\ell'}}}^2}{\mu_\ell + \mu_{\ell'}}. 
\end{equation}
Note that in principle \eqref{eq:dephasing-QFI} only holds true when $\{\ket{\psi_\ell}\}$ is a complete basis of $\mH_\mP^{\otimes N}$, that is, ${\rm span}\{\ket{\psi_\ell}\} = \mH_\mP^{\otimes N}$. However, we are allowed to restrict the summation in the RHS of \eqref{eq:dephasing-QFI} to states in the subspace $\mZ$, i.e. ${\rm span}\{\ket{\psi_\ell}\} = \mZ$, because $\Pi_\mZ \rho_\omega \Pi_\mZ =  \rho_\omega$ and $\Pi_\mZ \dot\rho_\omega \Pi_\mZ = \dot \rho_\omega$, i.e. any state perpendicular to $\mZ$ does not contribute to the QFI.

The derivative of $\rho_\omega$ w.r.t. $\omega$ is 
\begin{equation}
\begin{split}
\dot{\rho}_\omega
&= \frac{\partial \rho_\omega}{\partial p}{\dot p} + \frac{\partial \rho_\omega}{\partial \phi}{\dot \phi} = \sum_{j_1,\ldots,j_N=0}^1 \frac{\partial (1-p)^{(N - \sum_{k=1}^N j_k)}p^{(\sum_{k=1}^N j_k)}}{\partial \omega} \prod_{k=1}^N (\sigma_z^{(k)})^{j_k} \ket{\psi} \bra{\psi} \prod_{k=1}^N (\sigma_z^{(k)})^{j_k}\\
&\quad + \sum_{j_1,\ldots,j_N=0}^1  (1-p)^{(N - \sum_{k=1}^N j_k)}p^{(\sum_{k=1}^N j_k)} \prod_{k=1}^N (\sigma_z^{(k)})^{j_k} \frac{\partial \ket{\psi} \bra{\psi}}{\partial \omega} \prod_{k=1}^N (\sigma_z^{(k)})^{j_k}.
\end{split}
\end{equation}
Then we have 
\begin{equation}
\bra{\psi_\ell} \dot\rho_\omega \ket{\psi_{\ell'}} = a_{\ell\ell'} + i b_{\ell\ell'},
\end{equation}
where $a_{\ell\ell'} = \bra{\psi_\ell} \frac{\partial \rho_\omega}{\partial p}{\dot p} \ket{\psi_{\ell'}} \in \mathbb{R}$, $b_{\ell\ell'} = -i\bra{\psi_\ell} \frac{\partial \rho_\omega}{\partial \phi}{\dot \phi} \ket{\phi_{\ell'}} \in \mathbb{R}$. Therefore,
\begin{equation}
F(\rho_\omega) = 2 \sum_{\ell \ell':\mu_\ell + \mu_{\ell'} \neq 0} \frac{\abs{\bra{\psi_{\ell}}\dot{\rho}_\omega\ket{\psi_{\ell'}}}^2}{\mu_\ell + \mu_{\ell'}} = 2 \sum_{\ell \ell':\mu_\ell+\mu_{\ell'}\neq 0} \frac{\abs{a_{\ell\ell'}}^2 + \abs{b_{\ell\ell'}}^2}{\mu_{\ell} + \mu_{\ell'}} = F_{p}(\rho_\omega) + F_{\phi}(\rho_\omega),
\end{equation}
which is the same as \eqref{eq:dephasing-sum} in the main text.

\section{Optimal squeezed state for dephasing channels}
\label{app:dephasing-squeezed}

Let the input state $\ket{\psi_0} = e^{i\phi J_z}\ket{\psi_{\mu,\nu}}$, where $\ket{\psi_{\mu,\nu}}$ is an $N$-qubit spin-squeezed state
\begin{equation}
\ket{\psi_{\mu,\nu}} = e^{- i\nu J_x}e^{-\frac{i \mu}{2}J_z^2} e^{-i\frac{\pi}{2}J_y} \ket{0}^{\otimes N}.
\end{equation}
The output state is $\rho_\omega = \mD_\omega^{\otimes N} (\ket{\psi_0}\bra{\psi_0}) = (\mD_\omega|_{\phi = 0})^{\otimes N} (\ket{\psi}\bra{\psi})$. Then 
\begin{gather}
\braket{J_{x,y}}_{\rho_\omega} = (1-2p) \braket{J_{x,y}}_{\ket{\psi_{\mu,\nu}}},\\
\braket{J_{x,y}^2}_{\rho_\omega} = \frac{N}{4} + (1-2p)^2 \Big(\braket{J_{x,y}^2}_{\ket{\psi_{\mu,\nu}}} - \frac{N}{4}\Big),\\
\frac{\partial{\braket{J_x}_{\rho_\omega}}}{\partial p} \dot p = -2\dot p \braket{J_x}_{\ket{\psi_{\mu,\nu}}},\quad 
\frac{\partial{\braket{J_y}_{\rho_\omega}}}{\partial \phi} \dot \phi = (1-2p)\dot\phi \braket{J_x}_{\ket{\psi_{\mu,\nu}}}. 
\end{gather}
It was shown in \cite{kitagawa1993squeezed} that choosing $\nu = \frac{\pi}{2} - \frac{1}{2}\arctan \frac{b}{a}$, 
\begin{gather}
\braket{J_x}_{\ket{\psi_{\mu,\nu}}} = \frac{N}{2} \cos(\mu/2)^{N-1},\quad
\braket{J_y}_{\ket{\psi_{\mu,\nu}}} = 0,\\
\braket{\Delta J_x^2}_{\ket{\psi_{\mu,\nu}}} = \frac{N}{4} \left(N\left(1-\cos^{2(N-1)}\frac{\mu}{2}\right) - \left(\frac{N-1}{2}\right)a\right),\\
\braket{\Delta J_y^2}_{\ket{\psi_{\mu,\nu}}} =  \frac{N}{4} \left(1 + \frac{N-1}{4} \left(a - \sqrt{a^2+b^2}\right)\right),
\end{gather}
where 
$a = 1 - \cos^{N-2} \mu$, $b = 4 \sin\frac{\mu}{2}\cos^{N-2} \frac{\mu}{2}. $
Let $N\gg 1$, $\mu = \Theta(N^{-5/6})$, then
\begin{gather}
\braket{J_x}_{\ket{\psi_{\mu,\nu}}} \approx \frac{N}{2},\quad
\braket{\Delta J_x^2}_{\ket{\psi_{\mu,\nu}}} \approx O(N^{2/3}) ,\quad
\braket{\Delta J_y^2}_{\ket{\psi_{\mu,\nu}}} \approx O(N^{2/3}),
\end{gather}
and $\braket{\Delta J_x^2}_{\rho_\omega} \approx \braket{\Delta J_y^2}_{\rho_\omega} \approx p(1-p)N$, $\frac{\partial{\braket{J_x}_{\rho_\omega}}}{\partial p} \dot p \approx -\dot p N$ and $\frac{\partial{\braket{J_y}_{\rho_\omega}}}{\partial \phi} \dot \phi \approx (1-2p)\dot\phi N/2$.

\section{Optimizing the QFI when HNKS is violated}
\label{app:sql}

In this appendix, we optimize the QFI 
\begin{equation}
\frakF_\sql(\mD_{\tl,\omega}) = \frac{|\dot\xi|^2}{1 - |\xi|^2}
\end{equation}
{using \eqsref{eq:ef}{eq:xi}. }
We expand $T$ and $E_0E_1^\dagger$ around $\epsilon = 0$
\begin{gather}
T = e^{i\epsilon G} = 1 + i\epsilon G - \frac{\epsilon^2}{2}G^2 + O(\epsilon^3),\\
E_0E_1^\dagger = (1-\epsilon^2)EE^\dagger - \epsilon\sqrt{1-\epsilon^2}(EF^\dagger - FE^\dagger) - \epsilon^2 FF^\dagger = \sigma - i \epsilon \tilde\sigma - \epsilon^2 (FF^\dagger + EE^\dagger) + O(\epsilon^3),
\end{gather}
where $\sigma = EE^\dagger$ and $\tilde\sigma = i(FE^\dagger-EF^\dagger)$. Then 
\begin{gather}
\trace(T E_0E_1^\dagger) = 1 - 2 \epsilon^2 - \frac{\epsilon^2}{2} \trace(G^2 \sigma) - i \epsilon \trace(G \sigma) + \epsilon^2 \trace(G \tilde \sigma) + O(\epsilon^3),
\\
\trace(T (\dot{E}_0E_1^\dagger + E_0\dot{E}_1^\dagger)) =  i \epsilon \trace(G \dot\sigma) + O(\epsilon^2),
\end{gather}
where we used $\trace(F^\dagger F) = 1$ {and $\trace(\tilde \sigma) = 0$ because $\trace(E^\dagger F) = 0$.} Then 
\begin{align}
\frakF_\sql(\mD_{\tl,\omega}) 
&= \max_G \frac{\abs{\trace(G \dot\sigma)}^2 }{4 - 2 \trace(G \tilde \sigma) + \trace(G^2 \sigma) - \abs{\trace(G \sigma)}^2} + O(\epsilon)\\
&= \max_{G,x} \frac{\abs{\trace(G \dot\sigma)}^2 }{4 x^2 + 2 x \trace(G \tilde \sigma) + \trace(G^2 \sigma) - \abs{\trace(G \sigma)}^2} + O(\epsilon) \\
&= \max_{G} \frac{\abs{\trace(G \dot\sigma)}^2}{- \frac{\abs{\trace(G\tilde \sigma)}^2}{4} + \big( \trace(G^2 \sigma) - \abs{\trace(G \sigma)}^2 \big) } + O(\epsilon),
\label{eq:sql-G-app}
\end{align}
shown as \eqref{eq:sql-G} in the main text, where in the second line we used the fact that any rescaling of $G$ ($G \leftarrow -G/x$) should not change the optimal QFI. Note that to obtain the solution of the original $G$ in $T$, one need to rescale the final solution back using $G \leftarrow 4G/\trace(G\tilde\sigma)$. 

To find the optimal $G$, we first observe that $\trace(\dot\sigma) = \trace(\tilde\sigma) = 0$. Therefore, WLOG, we assume $\trace(G\sigma) = 0$ {because $G \leftarrow G - \trace(G)\frac{I}{r}$ does not change the target function}. Let the derivative of \eqref{eq:sql-G-app} be zero, we have
\begin{gather}
2\dot \sigma\Big(\trace(G^2  \sigma) - \frac{|\trace(G \tilde \sigma)|^2}{4}\Big) - \trace(G \dot \sigma) \Big((\sigma G + G\sigma) - \frac{2\trace(G\tilde \sigma)\tilde \sigma}{4}\Big) = 0,
\\\Leftrightarrow~~ \frac{\dot \sigma}{\trace(G \dot \sigma)}\Big(\trace(G^2  \sigma) - \frac{|\trace(G \tilde \sigma)|^2}{4}\Big) +  \frac{\trace(G\tilde \sigma)\tilde \sigma}{4} =  \frac{1}{2}(\sigma G + G\sigma),
\\\Leftrightarrow~~
G = L_\sigma[x\dot{\sigma} + y\tilde \sigma], \quad 4y = \trace(G\tilde\sigma) = \trace(L_\sigma[x\dot{\sigma} + y\tilde \sigma]\tilde \sigma),
\label{eq:sql-xy-1}
\\\Leftarrow~~
x = 4  - \trace(L_{\sigma}[\tilde \sigma]\tilde \sigma),\quad y = \trace(L_{\sigma}[\dot \sigma]\tilde \sigma).
\label{eq:sql-xy} 
\end{gather}
Note that in \eqref{eq:sql-xy-1} we used $x\dot\sigma+y\tilde\sigma = \frac{1}{2}(G\sigma + \sigma G)$ and $\trace(G^2\sigma) = \trace(G(x\dot\sigma+y\tilde\sigma))$.
Plug the optimal $G = L_\sigma[x\dot\sigma + y\tilde\sigma]$ into \eqref{eq:sql-G-app} where $x,y$ satisfies \eqref{eq:sql-xy}, we get 
\begin{equation}
\frakF_\sql(\mD_{\tl,\omega}) = \trace(L_\sigma[\dot\sigma]\dot{\sigma}) + \frac{\trace(L_\sigma[\dot\sigma]\tilde{\sigma})^2}{4 - \trace(L_\sigma[\tilde\sigma]\tilde{\sigma})}  + O(\epsilon) ,
\end{equation}
shown as \eqref{eq:sql-sigma} in the main text.

Next we express $\trace(L_\sigma[\dot\sigma]\dot{\sigma})$, $\trace(L_\sigma[\dot\sigma]\tilde{\sigma})$ and $\trace(L_\sigma[\tilde\sigma]\tilde{\sigma})$ in terms of $C$ and $\tilde C$. Let $\tau = E^\dagger E$, $\tilde\tau = E^\dagger F + F^\dagger E$, $\tau' = i E^\dagger \dot E -i \dot E^\dagger E$ such that
\begin{gather}
\tau_{ij} = \trace(C^\dagger K_i^\dagger K_j C),\;\tilde\tau_{ij} = \trace(\tilde C K_i^\dagger K_j),\\
\tau'_{ij} = i\trace(C^\dagger K_i^\dagger \dot K_j C) - i\trace(C^\dagger \dot K_i^\dagger  K_j C). 
\end{gather}
WLOG, assume $\tau_{ij} = \trace(C^\dagger K_i^\dagger K_j C) = \lambda_i \delta_{ij}$, which could always be achieved by performing a unitary transformation on $\vK$. We also have $\lambda_i > 0$ for all $i$ because $C$ is full rank and $\{\dket{K_i}\}_{i=1}^r$ are linearly independent. 
Using an orthonormal basis $\{\dket{i}\}_{i=1}^{d^2}$, where $\dket{i} = \frac{1}{\sqrt{\lambda_i}}\dket{K_i C}$ for $1 \leq i \leq r$. We have 
\begin{gather}
\sigma = 
\begin{pmatrix}
(\lambda_i \delta_{ij}) & 0 \\
0 & 0 
\end{pmatrix},\quad 
\dot\sigma = 
\begin{pmatrix}
\Big(\dbraket{K_iC|\dot{K}_j C}\sqrt{\frac{\lambda_j}{\lambda_i}} + \sqrt{\frac{\lambda_i}{\lambda_j}}  \dbraket{\dot{K}_iC|K_jC} \Big)& (\dbraket{\dot{K}_i C|j'} \sqrt{\lambda_i})\\
(\dbraket{i'|\dot{K}_j C} \sqrt{\lambda_j} )& 0 \\
\end{pmatrix}, \\
\tilde\sigma = 
\begin{pmatrix}
\Big(i \dbraket{K_iC|K_jD}\sqrt{\frac{\lambda_j}{\lambda_i}} - i \sqrt{\frac{\lambda_i}{\lambda_j}}  \dbraket{K_iD|K_jC} \Big)&( -i\dbraket{K_i D|j'} \sqrt{\lambda_i})\\
(i\dbraket{i'|K_j D} \sqrt{\lambda_j} )& 0 \\
\end{pmatrix},
\end{gather}
where $1\leq i,j \leq r$ and $r +1 \leq i',j' \leq d^2$. 
Then we can show \eqsref{eq:tau-1}{eq:tau-3} in the main text. 
\begin{equation}
\begin{split}
\trace(L_\sigma[\dot\sigma]\dot{\sigma}) 
&= 2 \sum_{i,j: \lambda_i + \lambda_j > 0} \frac{|(\dot\sigma)_{ij}|^2}{\lambda_i + \lambda_j} \\
&= 2 \sum_{i,j=1}^{r} \frac{\big|\dbraket{K_iC|\dot{K}_j C}\sqrt{\frac{\lambda_j}{\lambda_i}} + \sqrt{\frac{\lambda_i}{\lambda_j}}  \dbraket{\dot{K}_iC|K_jC}\big|^2}{\lambda_i + \lambda_j } + 4 \sum_{i'=r+1}^{d^2}\sum_{j=1}^{r} \frac{|\dbraket{i'|\dot{K}_j C} \sqrt{\lambda_j}|^2}{\lambda_j}\\
&= 4 \trace(C^\dagger \dot\vK^\dagger \dot\vK C) + 2 \sum_{i,j=1}^{r} \frac{\big|\dbraket{K_iC|\dot{K}_j C}\sqrt{\frac{\lambda_j}{\lambda_i}} + \sqrt{\frac{\lambda_i}{\lambda_j}}  \dbraket{\dot{K}_iC|K_jC}\big|^2}{\lambda_i + \lambda_j } - 2 \frac{|\dbraket{K_i C|\dot{K}_j C} |^2}{\lambda_i}\\
&= 4 \trace(C^\dagger \dot\vK^\dagger \dot\vK C) - 2 \sum_{i,j=1}^{r} \frac{|\tau_{ij}'|^2}{\lambda_i + \lambda_j} = 4 \trace(C^\dagger \dot\vK^\dagger \dot\vK C)  - \trace(L_\tau[\tau']\tau'),
\end{split}
\end{equation}
\begin{equation}
\begin{split}
\trace(L_\sigma[\tilde\sigma]\tilde\sigma) 
&= 2 \sum_{i,j:\lambda_i+\lambda_j > 0} \frac{\abs{(\tilde{\sigma})_{ij}}^2}{\lambda_i + \lambda_j}\\
&= 2 \sum_{i,j=1}^r \frac{\big|i \dbraket{K_iC|K_jD}\sqrt{\frac{\lambda_j}{\lambda_i}} - i \sqrt{\frac{\lambda_i}{\lambda_j}}  \dbraket{K_iD|K_jC}\big|^2}{\lambda_i + \lambda_j} + 4 \sum_{i'=r+1}^{d^2}\sum_{j=1}^r \frac{\abs{i\dbraket{i'|K_j D} \sqrt{\lambda_j} }^2}{\lambda_j}\\
&= 4 + 2 \sum_{i,j=1}^r \frac{\big|i \dbraket{K_iC|K_jD}\sqrt{\frac{\lambda_j}{\lambda_i}} - i \sqrt{\frac{\lambda_i}{\lambda_j}}  \dbraket{K_iD|K_jC}\big|^2}{\lambda_i + \lambda_j} - 2 \frac{|\dbraket{K_i C|K_j D} |^2}{\lambda_i}\\
&= 4 - 2 \sum_{ij} \frac{\abs{\tilde\tau_{ij}}^2}{\lambda_i + \lambda_j} = 4 - \trace(L_\tau[\tilde\tau]\tilde\tau),
\end{split}
\end{equation}
and 
\begin{equation}
\begin{split}
\trace(L_\sigma[\dot\sigma]\tilde{\sigma}) 
&= 2 \sum_{i,j:\lambda_i+\lambda_j\neq 0} \frac{\dot{\sigma}_{ij}\tilde{\sigma}_{ji}}{\lambda_i + \lambda_j}\\
&= 2 \sum_{i,j=1}^r \frac{\dot{\sigma}_{ij}\tilde{\sigma}_{ji}}{\lambda_i + \lambda_j} + 2 \sum_{i'=r+1}^{d^2}\sum_{j=1}^r \frac{\dot{\sigma}_{i'j}\tilde{\sigma}_{ji'}}{\lambda_j} + 2 \sum_{i'=r+1}^{d^2}\sum_{j=1}^r \frac{\dot{\sigma}_{ji'}\tilde{\sigma}_{i'j}}{\lambda_j}\\
&= - 2 \trace(\tilde C H) + 2 \sum_{i,j=1}^r \frac{\dot{\sigma}_{ij}\tilde{\sigma}_{ji}}{\lambda_i + \lambda_j} 
+ 2i \sum_{i,j=1}^r \frac{\dbraket{K_jD|K_iC}\dbraket{K_iC|\dot{K}_jC}}{\lambda_i} - \frac{\dbraket{\dot{K_j}C|K_iC}\dbraket{K_iC|K_jD}}{\lambda_i}\\
&= - 2 \trace(\tilde C H) + 2 \sum_{i,j=1}^r \frac{\tau_{ij}'\tilde\tau_{ji}}{\lambda_i + \lambda_j}   = - 2 \trace(\tilde C H) + \trace(L_\tau[\tau']\tilde\tau). 
\end{split}
\end{equation}
Therefore, we conclude that 
\begin{equation}
\frakF_\sql(\mD_{\tl,\omega}) \approx f(C,\tilde C) = 4 \trace(C^\dagger \dot\vK^\dagger \dot\vK C) - \trace(L_\tau[\tau']\tau')  + \frac{(- 2 \trace(\tilde C H)  + \trace(L_\tau[\tau']\tilde\tau))^2}{\trace(L_\tau[\tilde\tau]\tilde\tau)}.
\end{equation}

Next, we want to show
\begin{equation}
\label{eq:equiv}
\max_{\tilde C \in \bH_d} f(C,\tilde C) = \min_{h:\beta=0} 4\trace(C^\dagger \alpha C)
\end{equation}
when $C$ is full rank. To calculate the dual program of the RHS, we introduce a Hermitian matrix $\tilde C$ as a Lagrange multiplier of $\beta = 0$~\cite{boyd2004convex}. The Lagrange function is 
\begin{equation}
L(\tilde C,h) = 4 \trace(C^\dagger (\dot\vK-ih\vK)^\dagger(\dot\vK-ih\vK) C) + \trace(\tilde C (H + \vK^\dagger h \vK)),
\end{equation}
then
\begin{equation}
\begin{split}
\min_h L(\tilde C,h) 
&= \min_h 4 \trace(C^\dagger (\dot\vK-ih\vK)^\dagger(\dot\vK-ih\vK) C) + \trace(\tilde C (H + \vK^\dagger h \vK))\\
&= \min_h 4 \trace(C^\dagger \dot\vK^\dagger \dot\vK C) + 4\trace(\tau h^2) + 4\trace(iC^\dagger\vK^\dagger h\dot\vK C - iC^\dagger\dot\vK^\dagger h\vK C)+ \trace(\tilde C (H + \vK^\dagger h \vK))\\
&= \min_h 4 \trace(C^\dagger \dot\vK^\dagger \dot\vK C) + 4\trace(\tau h^2) + 4\trace(h^T \tau') + \trace(\tilde C H) + \trace(h^T \tilde\tau)\\
&=4 \trace(C^\dagger \dot\vK^\dagger \dot\vK C) + \trace(\tilde C H) - \frac{1}{8}\sum_{i,j=1}^r \frac{\big|4\tau'_{ij}+\tilde\tau_{ij}\big|^2}{\lambda_i+\lambda_j}.
\end{split}
\end{equation}
The dual program is 
\begin{equation}
\begin{split}
\max_{\tilde C}\min_h L(\tilde C,h) 
&= \max_{\tilde C} 4 \trace(C^\dagger \dot\vK^\dagger \dot\vK C) + \trace(\tilde C H) - \frac{1}{8}\sum_{i,j=1}^r \frac{16|\tau'_{ij}|^2 + |\tilde\tau _{ij}|^2 + 4(\tilde\tau _{ij}\tau'_{ji} + \tilde\tau _{ji}\tau'_{ij})}{\lambda_i+\lambda_j}\\
&= \max_{\tilde C,x} 4 \trace(C^\dagger \dot\vK^\dagger \dot\vK C) + x \trace(\tilde C H) - \frac{1}{8}\sum_{i,j=1}^r \frac{16|\tau'_{ij}|^2 + x^2|\tilde\tau _{ij}|^2 + 8x\tilde\tau _{ij}\tau'_{ji}}{\lambda_i+\lambda_j}\\
&=  \max_{\tilde C} 4 \trace(C^\dagger \dot\vK^\dagger \dot\vK C) - 2\sum_{i,j=1}^r \frac{|\tau'_{ij}|^2}{\lambda_i+\lambda_j} + \frac{\Big( - \trace(\tilde C H) + \sum_{i,j=1}^r \frac{\tilde\tau _{ij}\tau'_{ji}}{\lambda_i+\lambda_j} \Big)^2}{\frac{1}{2}\sum_{i,j=1}^r \frac{|\tilde\tau _{ij}|^2}{\lambda_i+\lambda_j}} = \max_{\tilde C} f(C,\tilde C),
\end{split}
\end{equation}
where we used the fact that $\tilde C \leftarrow x \tilde C$ does not change the result. \eqref{eq:equiv} is then proved.

Moreover, the optimal $\tilde C$ in \eqref{eq:equiv} must be traceless. Suppose $\tilde C$ is optimal in \eqref{eq:equiv}, we will prove that $\trace(\tilde C) = 0$. Let $z$ be a real number, 
\begin{equation}
q(z) := f(C,\tilde C + z C C^{\dagger}) = \frac{s(z)^2}{t(z)} + \text{const}. 
\end{equation}
Since $\max_z q(z) = q(0)$, we have $q'(0) = \frac{s(0)}{t(0)^2}\big(2s'(0)t(0) - s(0) t'(0)\big) = 0$. 
\begin{align}
s(z) &= - \trace((\tilde C+ z CC^\dagger) H) + \sum_{i,j=1}^r \frac{ (\tilde \tau_{ij} + z \lambda_i \delta_{ij}) \tau'_{ij}}{\lambda_i + \lambda_j},\\
s'(0) &= - \trace(CC^\dagger H) + \sum_{i=1}^r \frac{1}{2}\tau_{ii}' = 0,\\
t(z) &= \frac{1}{2}\sum_{i,j=1}^r \frac{\abs{\tilde \tau_{ij} + z \lambda_i \delta_{ij}}^2}{\lambda_i + \lambda_j} = \frac{1}{2}\sum_{i,j=1}^r \frac{\abs{\tilde \tau_{ij}}^2 + z \lambda_i \delta_{ij}(\tilde \tau_{ij}^* + \tilde \tau_{ij}) + z^2 \lambda_i^2 \delta_{ij}}{\lambda_i + \lambda_j},\\
t'(0) &= \frac{1}{2}\sum_{i,j=1}^r \frac{\lambda_i \delta_{ij}(\tilde \tau_{ij}^* + \tilde \tau_{ij}) 
}{\lambda_i + \lambda_j}
= \frac{1}{2}\sum_{i=1}^r \tilde\tau_{ii} = \frac{1}{2}\trace(\tilde C). 
\end{align}
Then $q'(0) = 0$ implies $\trace(\tilde C) = 0$.

\section{Numerical algorithm to find the optimal code when HNKS is violated}
\label{app:algo}

\subsection{Finding the optimal \texorpdfstring{$C$}{C}}
\label{app:algo-1}

We first describe a numerical algorithm finding a full rank $C^\diamond$ such that $\trace(C^{\diamond\dagger} C^\diamond) = 1$ and 
\begin{equation}
\min_{h:\beta = 0}4\trace(C^{\diamond\dagger} \alpha C^\diamond) > \frakF^{(u)}_\sql(\mE_\omega) - \eta/2.
\end{equation}
for any $\eta > 0$.
We first note that $\frakF^{(u)}_\sql(\mE_\omega) = \min_{h:\beta = 0} 4 \norm{\alpha}$ could be calculated via the following SDP~\cite{demkowicz2012elusive}, 
\begin{equation}
\label{eq:SDP}
\min_{h} x,\quad 
\text{subject to}~ 
\begin{pmatrix}
x I_d & \tilde K_1^\dagger & \cdots & \tilde K_r^\dagger \\
\tilde K_1 & I_{d'} & \cdots & 0 \\
\vdots & 0 & \ddots & \vdots \\
\tilde K_r & 0 & \cdots & I_{d'} 
\end{pmatrix} \succeq 0,\quad \beta = 0. 
\end{equation}
where $d$ and $d'$ are the input and output dimension of $\mE_\omega$, $I_{n}$ is a $n\times n$ identity matrix and $\tilde\vK = \dot \vK - ih\vK$. 

To find the full rank $C^\diamond$, we first find a density matrix $\rho^\diamond$ such that 
\begin{equation}
\min_{h:\beta = 0}4\trace(\rho^\diamond \alpha) = \min_{h:\beta = 0} 4 \norm{\alpha}. 
\end{equation}
It could be done via the following algorithm~\cite{zhou2019optimal}:
\begin{enumerate}[i),wide,labelwidth=!,labelindent=0pt]
\item Find an $h^\diamond$ using the SDP (\eqref{eq:SDP}), such that $\alpha^\diamond = \alpha|_{h=h^\diamond}$ satisfies $\norm{\alpha^\diamond} = \min_{h:\beta = 0} \norm{\alpha}$. 
\item Let $\Pi^\diamond$ be the projection onto the subspace spanned by all eigenstates corresponding to the largest eigenvalue of $\alpha^\diamond$, we find an optimal density matrix $\rho^\diamond$ satisfying $\Pi^\diamond \rho^\diamond \Pi^\diamond = \rho^\diamond$ and 
\begin{equation}
\Re[\trace(\rho^\diamond (i\vK^\dagger \Delta h) (\dot\vK - ih^\diamond\vK))] = 0,\quad \forall \eta \in \bH_r, ~\text{s.t.}~ \vK^\dagger \Delta h \vK = 0. 
\end{equation}
\end{enumerate}
Then $C^\diamond = \big((1-\eta') \rho^\diamond + \eta' \frac{I}{d}\big)^{1/2}$ where $\eta' = \eta/(2\frakF^{(u)}_\sql(\mE_\omega))$ is a full rank matrix satisfying 
\begin{equation}
\label{eq:eta-prime}
\min_{h:\beta = 0}4\trace(C^{\diamond\dagger} \alpha C^\diamond) > (1-\eta') \frakF^{(u)}_\sql(\mE_\omega) = \frakF^{(u)}_\sql(\mE_\omega) - \eta/2.
\end{equation}

The algorithm above could also be used to find $\rho^{\bdia}$ whose purification is the optimal input state of a single quantum channel $\mE_\omega$ achieving $\frakF_1(\mE_\omega)$: 
\begin{enumerate}[i),wide,labelwidth=!,labelindent=0pt]
\item Find an $h^{\bdia}$ using the SDP in \eqref{eq:SDP} without the requirement $\beta = 0$, such that $\alpha^{\bdia} = \alpha|_{h=h^{\bdia}}$ satisfies $\norm{\alpha^{\bdia}} = \min_{h} \norm{\alpha}$. 
\item Let $\Pi^{\bdia}$ be the projection onto the subspace spanned by all eigenstates corresponding to the largest eigenvalue of $\alpha^{\bdia}$, we find an optimal density matrix $\rho^{\bdia}$ satisfying $\Pi^{\bdia} \rho^{\bdia} \Pi^{\bdia} = \rho^{\bdia}$ and 
\begin{equation}
\Re[\trace(\rho^{\bdia} (i\vK^\dagger \Delta h) (\dot\vK - ih^{\bdia}\vK))] = 0,\quad \forall \Delta h \in \bH_r.
\end{equation}
\end{enumerate}

\subsection{Validity of the algorithm to find the optimal \texorpdfstring{$C$}{C}}

For completeness, we prove the validity of the above algorithm. 
According to Sion's minimax theorem~\cite{komiya1988elementary,do2001introduction}, for convex compact sets $\frakP \subset \bR^{m}$ and $\frakQ \subset \bR^{n}$ and $g:P\times Q \rightarrow \bR$ such that $g(x,y)$ is a continuous convex (concave) function in $x$ ($y$) for every fixed $y$ ($x$), then 
\begin{equation}
\label{eq:minimax}
\max_{y \in \frakQ}\min_{x \in \frakP} g(x,y) = \min_{x \in \frakP} \max_{y \in \frakQ} g(x,y).
\end{equation}

In particular, if $(x^\btri,y^\bdia)$ is a solution of $\max_{y \in \frakQ}\min_{x \in \frakP} g(x,y)$, then there must exists an $x^\bdia$ such that $(x^\bdia,y^\bdia)$ is a saddle point. 
Let $(x^\bdia,y^\btri)$ be a solution of $\min_{x \in \frakP} \max_{y \in \frakQ} g(x,y)$. Then we must have 
\begin{equation}
g(x^\btri,y^\bdia) \leq g(x^\bdia,y^\bdia) \leq g(x^\bdia,y^\btri).
\end{equation} 
According to \eqref{eq:minimax}, $g(x^\btri,y^\bdia) = g(x^\bdia,y^\btri)$ and all equalities must hold for the above equation. Moreover,
\begin{equation}
g(x^\bdia,y) \leq g(x^\bdia,y^\bdia) \leq g(x,y^\bdia),\quad \forall (x,y) \in \frakP\times\frakQ,
\end{equation} 
which means $(x^\bdia,y^\bdia)$ is a saddle point. 
For example, we can take $x = h \in \bH_r$, $y = CC^\dagger = \rho \in \frakS(\mH_\mP)$ and $g(x,y) = 4\trace(\rho \alpha)$. (We can also add the constraint $\beta = 0$ on $h$ which does not affect our discussion below). Then the solution of the above optimization problem is $\frakF_1(\mE_\omega)$ (or $\frakF_\sql(\mE_\omega)$ with the constraint $\beta = 0$). Note that we can always confine $h$ in a compact set such that the solutions are not altered and the minimax theorem is applicable~\cite{zhou2019optimal}. Let $(h^\btri,\rho^\bdia)$ be any solution of the optimization problem $\max_\rho \min_h 4 \trace(\rho\alpha)$. Then there exists an $h^\bdia$ such that $(h^\bdia,\rho^\bdia)$ is a saddle point. 
Similarly, if $g(x^\bdia,y^\btri)$ is a solution of $\min_{x \in \frakP} \max_{y \in \frakQ} g(x,y)$, which in our case is an SDP (\eqref{eq:SDP}). There must exists a $y^\bdia$ such that $(x^\bdia,y^\bdia)$ is a saddle point. Let $(h^\bdia,\rho^\btri)$ be any solution of the optimization problem $\min_h \max_\rho 4 \trace(\rho\alpha)$. Then there exists an $\rho^\bdia$ such that $(h^\bdia,\rho^\bdia)$ is a saddle point. Moreover, $(h^\bdia,\rho^\bdia)$ is a saddle point if and only if 
\begin{enumerate}[(i),wide,labelwidth=!,labelindent=0pt]
\item $\trace(\rho^\bdia \alpha^\bdia ) =  \norm{\alpha^\bdia}$, $~\Leftrightarrow~$ $\trace(\rho^\bdia \alpha^\bdia) \geq \trace(\rho \alpha^\bdia),\;\forall \rho$. 
\item $\Re[\trace(\rho^{\bdia} (i\vK^\dagger \Delta h) (\dot\vK - ih^{\bdia}\vK))] = 0,\;\forall \Delta h \in \bH_r,$  $~\Leftrightarrow~$ $\trace(\rho^\bdia \alpha^\bdia) \leq \trace(\rho^\bdia \alpha),\;\forall h$.
\end{enumerate}
It justifies the validity of the algorithm we described above.

\subsection{Finding the optimal \texorpdfstring{$\tilde C$}{tilde C}}

Next, we describe how to find $\tilde C^\diamond$ such that $
f(C^\diamond,\tilde C^\diamond) = \max_{\tilde C} f(C^\diamond,\tilde C) = \min_{h:\beta = 0}4\trace(C^{\diamond\dagger} \alpha C^\diamond)
$. According to \appref{app:sql}, 
\begin{equation}
f(C,\tilde C) = 4 \trace(C^\dagger \dot\vK^\dagger \dot\vK C) - 2\sum_{i,j=1}^r \frac{|\tau'_{ij}|^2}{\lambda_i+\lambda_j} + \frac{\Big( - \trace(\tilde C H) + \sum_{i,j=1}^r \frac{\tilde\tau _{ij}\tau'_{ji}}{\lambda_i+\lambda_j} \Big)^2}{\frac{1}{2}\sum_{i,j=1}^r \frac{|\tilde\tau _{ij}|^2}{\lambda_i+\lambda_j}},
\end{equation}
where we have assumed $\tau_{ij} = \trace(C^\dagger K_i^\dagger K_j C) = \lambda_i \delta_{ij}$. For a fixed $C$, $\tilde \tau$ is a linear function in $\tilde C$. We could always write 
\begin{equation}
f(C,\tilde C) = f_1(C) + \frac{|\dbraket{\tilde C|f_2(C)}|^2}{\dbraket{\tilde C|f_3(C)|\tilde C}},
\end{equation}
where $f_1(C)\in\bR$, $f_2(C)\in \bC^{d\times d}$ is Hermitian and $f_3(C) \in \bC^{d^2 \times d^2}$ is positive semidefinite. Moreover, $\dket{f_2(C)}$ is in the support of $f_3(C)$. $f_{1,2,3}(C)$ are functions of $C$ only. According to Cauchy-Schwarz inequality,
\begin{equation}
\max_{\tilde C} f(C,\tilde C) = f_1(C) + \dbraket{f_2(C)|f_3(C)^{-1}|f_2(C)},
\end{equation}
where the maximum is attained when $\dket{\tilde C} = f_3(C)^{-1}\dket{f_2(C)}$ and $^{-1}$ here means the Moore-Penrose pseudoinverse. Therefore, we take 
\begin{equation}
\dket{\tilde C^\diamond} = f_3(C^\diamond)^{-1}\dket{f_2(C^\diamond)}. 
\end{equation}

\section{An SDP to find the optimal input state of a noisy Mach-Zehnder interferometer}
\label{app:interfer}

Here we consider a two-arm Mach-Zehnder interferometer with one noisy arm and one noiseless arm, where the input state is an $M$-photon state
\begin{equation}
\label{eq:boson-input}
\ket{\psi_0} = \sum_{m=0}^{M} \gamma_m \ket{m}\ket{M-m}. 
\end{equation}
Here $\ket{m,M-m}$ represents a two-mode Fock state where $m$ is the number of photons in the first mode and $M-m$ is the number of photons in the second. The noisy quantum channel $\mE_\omega(\cdot)$ acting on the first mode is described by the Kraus operators
\begin{equation}
K_i = \sqrt{\frac{(1-p)^{i}}{i!}} e^{-i\omega \ann^\dagger \ann} p^{\frac{1}{2}{\ann^\dagger \ann}} \ann^i,\quad i=0,1,\ldots,M, 
\end{equation}
where $\omega$ is the unknown phase to be estimated, $\ann$ is the photon annihilation operator, $0 < p < 1$ is the loss rate and $K_i$ is associated with losing $i$ photons. 
Note that we are allowed to truncate the maximum photon number at $M$ because of the restriction on the input state~(\eqref{eq:boson-input}). 

We will show that the algorithm described in \appref{app:algo} naturally gives an SDP for obtaining the optimal $\{\gamma_{m}\}_{m=0}^M$. We emphasize here that it was already shown in \cite{demkowicz2009quantum} that solving for the optimal input state in an interferometer with two noisy arms is a convex optimization problem. Here we provide an alternative algorithm as a demonstration of our approach which also contains a proof that states of the form \eqref{eq:boson-input} are optimal for $\mE_\omega$. 

Recall that given $\mE_\omega$, we can find an optimal input state achieving $\frakF_1(\mE_\omega)$ by purifying $\rho^\bdia$ which is a solution of 
\begin{equation}
\frakF_1(\mE_\omega) = \max_\rho \min_h 4 \trace(\rho\alpha) = \min_h 4 \norm{\alpha}. 
\end{equation}
We show that the optimization problem above has a diagonal solution of $\rho$. Note that 
\begin{equation}
\begin{split}
\alpha &= \sum_{i=0}^M \left(\dot K_i - i \sum_{i,j=0}^M h_{ij} K_j\right)^\dagger \left(\dot K_i - i \sum_{j'=0}^M h_{ij'} K_{j'}\right) \\
&= \sum_{i=0}^M  \left(\dot K_i - i h_{ii} K_i\right)^\dagger \left(\dot K_i - i h_{ii} K_{i}\right) + \sum_{i=0}^M\sum_{j\neq i} K_j^\dagger h^*_{ij} h_{ij} K_j + \text{off-diagonal terms},
\end{split}
\end{equation}
where we divided $\alpha$ into diagonal terms and off-diagonal terms (in the Fork basis). The second term is always non-negative and the off-diagonal terms will only increase $\norm{\alpha}$. It is then clear that we can always assume the optimal $h^\bdia$ and the corresponding $\alpha^\bdia$ are diagonal because
\begin{equation}
\norm{\alpha} \geq \Big\| \sum_{i=0}^M \big(\dot K_i - i h_{ii} K_i\big)^\dagger  \big(\dot K_i - i h_{ii} K_{j}\big) \Big\|.
\end{equation}
Choose a diagonal $h^\bdia$ and let $\Pi^{\bdia}$ be the projection onto the subspace spanned by all eigenstates corresponding to the largest eigenvalue of $\alpha^{\bdia}$, $\rho^{\bdia}$ is optimal if it satisfies $\Pi^{\bdia} \rho^{\bdia} \Pi^{\bdia} = \rho^{\bdia}$ and 
\begin{equation}
\Re[\trace(\rho^{\bdia} (i\vK^\dagger \Delta h) (\dot\vK - ih^{\bdia}\vK))] = 0,\quad \forall \Delta h \in \bH_{M+1}.
\end{equation}
We observe that when $\rho^\bdia$ is optimal, the equation above still holds by replacing $\rho^\bdia$ with its diagonal part. Then any diagonal $\rho^\bdia$ which satisfies
\begin{equation}
\Re[\trace(\rho^{\bdia} i K_i^\dagger \Delta h_{ii} (\dot K_i - ih^{\bdia}_{ii} K_i))] = 0,\quad \forall\{\Delta h_{ii}\}_{i=0}^M \in \bR^{M+1}
\end{equation}
is optimal. Therefore, we can always assume the input state has the form \eqref{eq:boson-input}. Moreover, by assuming $h$ and $\rho$ are diagonal, we only need to deal with diagonal operators in this algorithm and the problem is essentially a quadratically constrained quadratic program, which might admit more efficient numerical methods than the SDP formulation.

\section{Channel QFIs for the single-qubit depolarizing channels}
\label{app:depolarizing}

Here we calculate $\frakF_1$, $\frakF_\sql$ and $\frakF_\hl$ for depolarizing channels
\begin{equation}
\begin{split}
\mN^{{\rm d}}_{\omega}(\rho) &= (1-p) e^{-\frac{i\omega}{2}\sigma_z}\rho e^{\frac{i\omega}{2}\sigma_z} +  p_x \sigma_x e^{-\frac{i\omega}{2}\sigma_z}\rho e^{\frac{i\omega}{2}\sigma_z}  \sigma_x  \\ &\qquad\quad  + p_y \sigma_y e^{-\frac{i\omega}{2}\sigma_z}\rho e^{\frac{i\omega}{2}\sigma_z}  \sigma_y + p_z \sigma_z e^{-\frac{i\omega}{2}\sigma_z}\rho e^{\frac{i\omega}{2}\sigma_z}  \sigma_z = \sum_{i=1}^4 K_i \rho K_i^\dagger, 
\end{split}
\end{equation}
where $K_1 = \sqrt{1-p} e^{- \frac{i \omega}{2} \sigma_z}, K_2 = \sqrt{p_x} \sigma_x e^{- \frac{i \omega}{2} \sigma_z}, K_3 = \sqrt{p_y} \sigma_y e^{- \frac{i \omega}{2} \sigma_z}, K_4 = \sqrt{p_z} \sigma_z e^{- \frac{i \omega}{2} \sigma_z}$. 
\begin{gather}
\vK = 
\begin{pmatrix}
\sqrt{1-p}\\
\sqrt{p_x} \sigma_x \\
\sqrt{p_y} \sigma_y \\
\sqrt{p_z} \sigma_z \\
\end{pmatrix} e^{- \frac{i \omega}{2} \sigma_z}
,\quad 
\dot\vK = 
\begin{pmatrix}
-\frac{i}{2} \sqrt{1-p} \sigma_z \\
-\frac{1}{2} \sqrt{p_x} \sigma_y \\
\frac{1}{2} \sqrt{p_y} \sigma_x \\
-\frac{i}{2}\sqrt{p_z} \\
\end{pmatrix}e^{- \frac{i \omega}{2} \sigma_z}
,\\
\beta = i\vK^\dagger (\dot\vK - ih\vK) = \frac{1}{2}\sigma_z + \vK^\dagger h \vK.
\end{gather}

\begin{equation}
\beta = 0 ~~~\Rightarrow~~~ 
\begin{array}{c}
(1-p)h_{11} + p_x h_{22} + p_y h_{33} + p_z h_{44} = 0,\\
\sqrt{(1-p)p_x}(h_{12} + h_{21}) + i \sqrt{p_y p_z} h_{34} - i \sqrt{p_y p_z} h_{43} = 0,\\
\sqrt{(1-p)p_y}(h_{13} + h_{31}) - i \sqrt{p_x p_z} h_{24} + i \sqrt{p_x p_z} h_{42} = 0,\\
\frac{1}{2} + \sqrt{(1-p)p_z}(h_{14} + h_{41}) + i \sqrt{p_x p_y} h_{23} - i \sqrt{p_x p_y} h_{32} = 0. \\
\end{array}
\end{equation}
Clearly, HNKS is satisfied if and only if $p_x = p_z = 0$ or $p_y = p_z = 0$. 
It is easy to see that when $h_{ij} = 0$ for all $i,j$ except $h_{23}$, $h_{32}$, $h_{14}$ and $h_{41}$, $\alpha = \norm{\alpha} I$, $\norm{\alpha}$ takes its minimum and 
\begin{equation}
\norm{\alpha} = \frac{1}{4} + \sqrt{(1-p)p_z} (h_{14} + h_{41}) + i\sqrt{p_xp_y} (h_{23}-h_{32}) + (1-p+p_z)\abs{h_{14}}^2 + (p_x+p_y) \abs{h_{23}}^2 
\end{equation}
Then 
\begin{equation}
\frakF_1(\mN^{{\rm d}}_\omega) = 4 \min_{h} \norm{\alpha} = 1 - 4 \left(\frac{p_xp_y}{p_x+p_y} + \frac{(1-p)p_z}{1-p+p_z}\right).
\end{equation}
When HNKS is satisfied,
\begin{equation}
\frakF_\hl(\mN^{{\rm d}}_\omega) = 4 \min_{h} \norm{\beta}^2  = 1,
\end{equation}
and when HNKS is violated, 
\begin{equation}
\frakF_\sql(\mN^{{\rm d}}_\omega) = 4 \min_{h:\beta = 0} \norm{\alpha} = - 1 + \frac{1}{4}\left(\frac{p_xp_y}{p_x+p_y} + \frac{(1-p)p_z}{1-p+p_z}\right)^{-1}.
\end{equation}

\section{Solving the optimal QEC code for amplitude damping channels}
\label{app:ad}

In this appendix, we use the algorithm in \appref{app:algo} to solve for the optimal QEC protocol analytically for amplitude damping channels with two Kraus operators: 
\begin{equation}
\vK = \begin{pmatrix}
\ket{0}\bra{0} + \sqrt{1 - p}\ket{1}\bra{1} \\ 
\sqrt{p}\ket{0}\bra{1}
\end{pmatrix}e^{-i\frac{\omega}{2}\sigma_z} = 
\begin{pmatrix}
\ket{0}\bra{0}e^{-i\frac{\omega}{2}} + \sqrt{1 - p} \ket{1}\bra{1}e^{i\frac{\omega}{2}}\\
\sqrt{p} \ket{0}\bra{1}e^{i\frac{\omega}{2}}
\end{pmatrix}.
\end{equation}
Clearly, $H = i \vK^\dagger \dot\vK = \sigma_z/2$. 

\subsection{Finding the optimal \texorpdfstring{$C$}{C}}

First, we want to find a full rank normalized $C^\diamond$, such that $\min_{h:\beta = 0}4\trace(C^{\diamond\dagger} \alpha C^\diamond)$ is close to $\frakF_{\sql}^{(u)}(\mE_\omega)$ and we will follow the algorithm described in \appref{app:algo-1}.


We first compute $\alpha$ and $\beta$. Note that  
\begin{equation}
\dot\vK 
= 
\begin{pmatrix}
-\frac{i}{2}\ket{0}\bra{0} + \frac{i}{2}\sqrt{1-p} \ket{1}\bra{1}\\
\frac{i}{2}\sqrt{p} \ket{0}\bra{1}
\end{pmatrix}
e^{-i\frac{\omega}{2}\sigma_z}.
\end{equation}
We first observe that in order to make $\beta = i \vK^\dagger (\dot\vK - i h \vK) = 0$, $h$ has to be diagonal and then 
\begin{equation}
\dot\vK - ih\vK
= 
\begin{pmatrix}
(-\frac{i}{2} - i h_{11})\ket{0}\bra{0} + (\frac{i}{2} - i h_{11})\sqrt{1-p} \ket{1}\bra{1}\\
(\frac{i}{2} - i h_{22})\sqrt{p} \ket{0}\bra{1}
\end{pmatrix}
e^{-i\frac{\omega}{2}\sigma_z}. 
\end{equation}
We will also assume $\omega = 0$ for simplicity. 
\begin{gather}
\beta = i \vK^\dagger (\dot\vK - i h \vK) =  \left(\frac{1}{2}+h_{11}\right)\ket{0}\bra{0} + \left(-\frac{1}{2}+h_{11}\right)(1-p) \ket{1}\bra{1} + \left(-\frac{1}{2}+h_{22}\right)p\ket{1}\bra{1} = 0, \\
~~~\Rightarrow~~~
h_{11} = - \frac{1}{2},  
\quad 
h_{22} = \frac{2 - p}{2 p}. 
\end{gather}
Therefore 
$
\alpha = (\dot\vK - i h \vK)^\dagger (\dot\vK - i h \vK) = \left((1-p) + \frac{(1-p)^2}{p}\right) \ket{1}\bra{1} = \frac{1-p}{p} \ket{1}\bra{1}$. 

Since there is only one solution of $h$ such that $\beta = 0$, there is no need to solve $\min_{h:\beta = 0}\norm{\alpha}$ using an SDP and the only solution is: $\alpha^\diamond = \frac{1-p}{p} \ket{1}\bra{1}$ and $\frakF_\sql(\mN^{{\rm ad}}_\omega) = 4(1-p)/p$. We could take $C^\diamond =  \sin\delta \ket{0}\bra{0} + \cos\delta \ket{1}\bra{1}$ where ${\delta}$ is small. Note that here we use the small constant $\delta$ instead of $\eta'$ in \eqref{eq:eta-prime} for simplicity. They are related by $\eta'/2 = \sin^2(\delta)$. 


\subsection{Finding the optimal \texorpdfstring{$\tilde C$}{tilde C}}

Next we find the optimal $\tilde C^\diamond$ which minimizes 
\begin{equation}
\frac{\Big( - \trace(\tilde C H) + \sum_{i,j=1}^r \frac{\tilde\tau _{ij}\tau'_{ji}}{\lambda_i+\lambda_j} \Big)^2}{\frac{1}{2}\sum_{i,j=1}^r \frac{|\tilde\tau _{ij}|^2}{\lambda_i+\lambda_j}} = \frac{|\dbraket{\tilde C | f_2(C)}|^2}{\dbra{\tilde C}f_3(C)\dket{\tilde C}},
\end{equation}
and the solution is $\dket{\tilde C^\diamond} = f_3(C^\diamond)^{-1}\dket{f_2(C^\diamond)}$.

We first compute 
\begin{gather}
\tau= 
\begin{pmatrix}
\sin^2\delta + (1-p)\cos^2\delta & 0 \\
0 & p \cos^2\delta
\end{pmatrix}
\approx
\begin{pmatrix}
1 - p & 0 \\
0 & p
\end{pmatrix}
,\\
\tau'=
\begin{pmatrix}
\sin^2\delta - (1-p)\cos^2\delta & 0 \\ 
0 & -p \cos^2\delta\\
\end{pmatrix}
\approx
\begin{pmatrix}
-(1-p) & 0 \\ 
0 & -p\\
\end{pmatrix}, \\
\tilde\tau = 
\begin{pmatrix}
\trace(\tilde C \begin{pmatrix}1 & 0 \\ 0 & 1 - p \end{pmatrix}) 
& \trace(\tilde C \begin{pmatrix}0 & \sqrt{p} \\ 0 & 0 \end{pmatrix}) 
\\ 
\trace(\tilde C \begin{pmatrix}0 & 0 \\ \sqrt{p} & 0 \end{pmatrix}) 
& \trace(\tilde C \begin{pmatrix}0 & 0 \\ 0 & p \end{pmatrix})\\
\end{pmatrix}, 
\end{gather}
where by ``$\approx$'' we ignore the small contribution of $O(\delta)$. 
Then we have 
\begin{gather}
f_2(C^\diamond) \approx - \ket{00}, \\
f_3(C^\diamond)^{-1} \approx p\ket{01}\bra{01} + p\ket{10}\bra{10} + \qquad\qquad\qquad\qquad\qquad\qquad\qquad  \nonumber\\ \qquad\qquad\qquad\qquad\qquad\qquad\qquad (\ket{00}+(1-p)\ket{11})(\bra{00}+(1-p)\bra{11}) \frac{1}{2(1-p)} 
+ (p\ket{11})(p\bra{11})  \frac{1}{2p},
 \\ 
f_3(C^\diamond)^{-1} \approx \frac{1}{p}\ket{01}\bra{01} + \frac{1}{p}\ket{10}\bra{10}
+ \frac{2}{p}((1-p)\ket{00}\bra{00} + \ket{11}\bra{11} 
- (1-p)\ket{00}\bra{11} - (1-p)\ket{11}\bra{00}). 
\end{gather}
Then $\dket{\tilde C^\diamond} \approx \ket{00}-\ket{11}$ and we could take $D^\diamond = \cos\delta\ket{0}\bra{0} - \sin\delta\ket{1}\bra{1}$.

\subsection{Attaining the asymptotic QFI}

Now we have the optimal code from the previous two steps: 
\begin{equation}
\ket{0_\tl} 
= \sin(\delta + \epsilon)\ket{0}_\mP \ket{00}_\mA + \cos(\delta + \epsilon)\ket{1}_\mP \ket{10}_\mA,\quad 
\ket{1_\tl}  
= \sin(\delta - \epsilon)\ket{0}_\mP \ket{01}_\mA + \cos(\delta - \epsilon)\ket{1}_\mP \ket{11}_\mA. 
\end{equation}
where $\delta$ and $\epsilon = o(\delta)$ are small values. 
The last step is to find the exact relation between $\delta$ and $\epsilon$ and $\frakF_\sql(\mN^{\rm ad}_{\omega}) - \frakF_\sql(\mD_{\tl,\omega}) $. 

To do so, we need the near-optimal recovery channel computed using \eqref{eq:sql-G-opt}: 
\begin{equation}
G_{\text{opt}} = \frac{2i}{\sqrt{1-p}}\ket{00}\bra{11} + \frac{-2i}{\sqrt{1-p}}\ket{11}\bra{00} := 
\begin{pmatrix}
0 & 0 & 0 & \frac{2i}{\sqrt{1-p}} \\
0 & 0 & 0 & 0 \\
0 & 0 & 0 & 0 \\
\frac{-2i}{\sqrt{1-p}} & 0 & 0 & 0 \\
\end{pmatrix},
\end{equation}
and 
\begin{equation}
T_{\text{opt}} = e^{i\epsilon G_{\text{opt}}} = 
\begin{pmatrix}
\cos\left(\frac{2\epsilon}{\sqrt{1-p}}\right) & 0 & 0 & -\sin\left(\frac{2\epsilon}{\sqrt{1-p}}\right) \\
0 & 1 & 0 & 0 \\
0 & 0 & 1 & 0 \\
\sin\left(\frac{2\epsilon}{\sqrt{1-p}}\right) & 0 & 0 & \cos\left(\frac{2\epsilon}{\sqrt{1-p}}\right)
\end{pmatrix}. 
\end{equation}
Then using \eqref{eq:xi}, and 
\begin{equation}
E_{0,1} = 
\begin{pmatrix}
\dket{K_1 A_{0,1}} & \dket{K_2 A_{0,1}}
\end{pmatrix} =
\begin{pmatrix}
\sin(\delta\pm\epsilon) & 0 \\
0 & \sqrt{p}\cos(\delta\pm\epsilon) \\
0 & 0 \\
\sqrt{1-p}\cos(\delta\pm\epsilon) & 0\\
\end{pmatrix}, 
\end{equation}
we finally get 
\begin{equation}
\begin{split}
\xi &=  p (\cos (2 \delta )+\cos (2 \epsilon )) \sin ^2\left(\frac{\epsilon }{\sqrt{1-p}}\right)  \\  &\qquad\qquad+\sqrt{1-p} \sin (2 \epsilon ) \sin \left(\frac{2 \epsilon }{\sqrt{1-p}}\right) +\cos (2 \epsilon ) \cos \left(\frac{2 \epsilon }{\sqrt{1-p}}\right) 
 = 1 - \frac{2p\sin^2\delta}{1-p} \epsilon^2 + O(\epsilon^4), 
\end{split}
\end{equation}
\begin{equation}
\dot\xi = -i \sqrt{1-p} \sin (2 \delta ) \sin \left(\frac{2 \epsilon }{\sqrt{1-p}}\right) = -2i\sin(2\delta)\epsilon + O(\epsilon^3). 
\end{equation}
Clearly, 
\begin{equation}
\frakF_\sql(\mD_{\tl,\omega}) = \frac{4(1-p)\cos^2\delta}{p} + O(\epsilon^2),
\end{equation}
as expected. When $\delta$ is small and $\epsilon = o(\delta)$, we would have $\frakF_\sql(\mD_{\tl,\omega}) \approx \frakF_\sql(\mN^{{\rm ad}}_\omega) $.

\end{document}